\documentclass[fleqn,usenatbib]{mnras}

\usepackage{newtxtext}
\usepackage{newtxmath}

\usepackage{natbib}
\usepackage[T1]{fontenc}
\usepackage{aecompl}

\usepackage{graphicx}							
\usepackage{bm}									
\graphicspath{{graphics/}}					
\usepackage{amsmath}							
\usepackage{amssymb}							
\usepackage{physics}							
\usepackage[nameinlink, capitalize]{cleveref}	
\usepackage{hyperref}
\usepackage{array}
\usepackage{xcolor}

\newcommand{\zl}{z_\mathrm{l}}
\newcommand{\zs}{z_\mathrm{s}}
\newcommand{\Msun}{M_\odot}

\newcommand{\Msub}{M_\mathrm{sub}}
\newcommand{\Mhm}{M_\mathrm{hm}}
\newcommand{\fsub}{f_\mathrm{sub}}
\newcommand{\Menc}{M_{2\erad}}

\newcommand{\etheta}{\theta_\varepsilon}
\newcommand{\Ml}{M_\mathrm{l}}
\newcommand{\erad}{\theta_\mathrm{E}}
\newcommand{\sigmav}{\sigma_V}
\newcommand{\reff}{r_\mathrm{eff_l}}
\newcommand{\Dls}{D_\mathrm{ls}}

\newcommand{\Dl}{D_\mathrm{l}}

\newcommand{\lensx}{\theta_{\mathrm{l}_x}}
\newcommand{\lensy}{\theta_{\mathrm{l}_y}}

\newcommand{\prob}[2]{\mathrm{Pr}\left(#1\vert#2\right)}	

\newcommand{\vmax}{v_\mathrm{max}}

\newcommand{\rmax}{r_\mathrm{max}}
\newcommand{\Mmax}{M_\mathrm{max}}

\newcommand{\rs}{r_\mathrm{s}}
\newcommand{\rhos}{\rho_\mathrm{s}}

\newcommand{\detsig}{s_\mathrm{sub}}
\newcommand{\musub}{\mu_\mathrm{sub}}
\newcommand{\mmin}{m_0}
\newcommand{\mmax}{m_1}

\defcitealias{Collett2015}{C15}

\title[Subhalo sensitivity in Euclid]{Sensitivity of strong lensing observations to dark matter substructure: a case study with Euclid}

\author[C. M. O'Riordan et al.]{
Conor M. O'Riordan,$^{1}$\thanks{E-mail: conor@mpa-garching.mpg.de} Giulia Despali$^{2,3}$, Simona Vegetti$^{1}$,\newauthor\,Mark R. Lovell$^{4}$ \& \'Angeles Molin\'e$^{5}$
	\\
	$^{1}$Max Planck Institut f\"{u}r Astrophysik, Karl-Schwarzschild-Stra{\ss}e 1, 85748 Garching bei M{\"u}nchen, Germany\\
	$^{2}$Institut f\"{u}r Theoretische Astrophysik, Zentrum f\"{u}r Astronomie, Heidelberg Universit\"{a}t, Albert-Ueberle-Str. 2, 69120, Heidelberg, Germany\\
	$^{3}$Dipartimento di Fisica e Astronomia "A. Righi", Alma Mater Studiorum Universit\'a di Bologna, Via Piero Gobetti 93/2, I-40129 Bologna, Italy\\
	$^{4}$Center for Astrophysics and Cosmology, Science Institute, University of Iceland, Dunhagi 5, 107 Reykjavik, Iceland\\
	$^{5}$Departamento de F\'isica, ETSI Navales, Universidad Polit\'ecnica de Madrid, Avda. de la Memoria, 4, 28040 Madrid, Spain\\
}
\date{Accepted XXX. Received YYY; in original form ZZZ}
\pubyear{2023}

\begin{document}
\label{firstpage}
\pagerange{\pageref{firstpage}--\pageref{lastpage}}
\maketitle

\begin{abstract}
	We introduce a machine learning method for estimating the sensitivity of strong lens observations to dark matter subhaloes in the lens. Our training data include elliptical power-law lenses, Hubble Deep Field sources, external shear, and noise and PSF for the Euclid VIS instrument. We set the concentration of the subhaloes using a $v_\mathrm{max}$-$r_\mathrm{max}$ relation. We then estimate the dark matter subhalo sensitivity in $16{,}000$ simulated strong lens observations with depth and resolution resembling Euclid VIS images. We find that, with a $3\sigma$ detection threshold, $2.35$ per cent of pixels inside twice the Einstein radius are sensitive to subhaloes with a mass $M_\mathrm{max}\leq 10^{10}M_\odot$, $0.03$ per cent are sensitive to $M_\mathrm{max}\leq 10^{9}M_\odot$, and, the limit of sensitivity is found to be $M_\mathrm{max}=10^{8.8\pm0.2}M_\odot$. Using our sensitivity maps and assuming CDM, we estimate that Euclid-like lenses will yield $1.43^{+0.14}_{-0.11}[f_\mathrm{sub}^{-1}]$ detectable subhaloes per lens in the entire sample, but this increases to $35.6^{+0.9}_{-0.9}[f_\mathrm{sub}^{-1}]$ per lens in the most sensitive lenses. Estimates are given in units of the inverse of the substructure mass fraction $f_\mathrm{sub}^{-1}$. Assuming $f_\mathrm{sub}=0.01$, one in every $70$ lenses in general should yield a detection, or one in every $\sim$ three lenses in the most sensitive sample. From $170,000$ new strong lenses detected by Euclid, we expect $\sim 2500$ new subhalo detections. We find that the expected number of detectable subhaloes in warm dark matter models only changes relative to cold dark matter for models which have already been ruled out, i.e., those with half-mode masses $M_\mathrm{hm}>10^8M_\odot$.
		
\end{abstract}



\section{Introduction}

In cosmological models involving dark matter, galaxies reside in dark matter superstructures called haloes. Numerical simulations show that these haloes form hierarchically, merging with their neighbours and agglomerating smaller haloes \citep{Springel2008}. The distribution of these subhaloes in mass, called the subhalo mass function (SHMF) depends on the free-streaming properties of the dark matter, which is typically parametrised as a thermal relic dark matter particle mass, $m_\rmn{TR}$. In the canonical model, cold dark matter (CDM), the mass function is scale free and the number of objects at a given mass scales inversely with mass. For warmer models, i.e., smaller $m_\rmn{TR}$, we expect a suppression in dark matter structure formation for objects below a certain mass called the half mode mass, $\Mhm$. Therefore, measuring the number and mass of dark matter subhaloes in galaxies provides a constraint on the dark matter model $m_\rmn{TR}$.

Strong gravitational lensing provides one method for measuring the distribution of dark matter subhaloes in the universe, amongst others \citep[e.g.][]{Koopmans2005,Dalal2002,Mao1998}. Subhaloes in the proximity of lensed images have a miniscule, but measurable effect on the local magnification. For point source objects like quasars, the effect of the subhalo is to produce anomalous flux ratios in the lensed images \citep{Xu2015,Bradac2002}. When the source is an extended object, a technique called gravitational imaging is used \citep{Galan2022, Vernardos2022,Vegetti2009}. In gravitational imaging, small corrections to the potential, beyond that of the smooth lens galaxy, are found that improve the fit to the lensed images. 
The density field for this corrected potential can then reveal the locations and masses of subhaloes. Alternatively, the presence of substructures can be described analytically \citep[e.g.][]{He2022, Dylan2018}, or via the power-spectrum of mass density fluctuations \citep[e.g.][]{Chatterjee2018}. So far, a small number of dark matter substructures have been detected \citep{Hezaveh2016,Vegetti2012,Vegetti2010}. In any CDM or WDM model, the number of detectable subhaloes expected in a typical lens with current instrumentation is of order unity, and so non-detections  of subhaloes are common \citep{Nightingale2022,Ritondale2019,Vegetti2014}. It is therefore crucial to quantify the limits in subhalo mass of any non-detection in order to constrain the subhalo mass function.

Traditionally, the ability of a strong lens observation to detect subhaloes is quantified via the sensitivity function. In this procedure, the strong lens is modelled both with and without a subhalo in every pixel, over a range of subhalo masses \citep{Ritondale2019,Vegetti2014}. Bayesian model comparison then gives the difference in the log-evidence, $\Delta \log\varepsilon$ between the models with and without subhaloes as a function of subhalo mass and position. By defining a threshold in $\Delta \log\varepsilon$ at which a detection would be acceptable, one finds the minimum detectable subhalo mass in every pixel.

The nature of evidence calculations and the complexity of strong lens modelling mean that the sensitivity function is expensive to compute, often taking hundreds or thousands of CPU hours for a single observation. Calculating the sensitivity function is in fact the most expensive component of a gravitational imaging study. This is not necessarily an issue when the number of lenses available for study is small, as has been the case in the past. However, automations in lens finding and the advent of the sky-survey era means the number of known lenses is growing rapidly, and will continue to do so in the near future. The Euclid survey alone is predicted to yield more than $10^5$ new strong lenses, alongside similarly impressive contributions from the Dark Energy Survey (DES) and the Vera Rubin Observatory \citep[][hereafter \citetalias{Collett2015}]{Collett2015}. It would be infeasible to conduct gravitational imaging studies on such a large number of objects with the current method, and so, we are motivated to find a more efficient method for calculating the sensitivity function.

In this paper, we demonstrate a new method for calculating the sensitivity to dark matter subhaloes in strong lens observations. Our method relies on machine learning, which is now widely used in strong lensing. This is especially true in lens finding, where convolutional neural networks (CNNs) are a natural choice \citep{Wilde2022,Shu2022,Lanusse2018}. Machine learning has also been used to estimate parameters for lens models \citep{Gu2022,Schuldt2021,Chianese2020,Hezaveh2017}. In our area of interest, detecting dark matter substructure, ML is also proving to be useful for replacing all or part of the traditional gravitational imaging pipeline. For example: \citet{Vernardos2020} show that a CNN can reliably estimate potential corrections in the lens; \citet{Wagner-Carena2022} use simulation-based inference to estimate the SHMF in a population of lenses; \citet{Ostdiek2022} and \citet{DiazRivero2020} show that direct detections with ML are possible in mock observations; and \citet{Coogan2020} develop a source light and lens potential modelling tool using ML.

Our approach for estimating subhalo sensitivity consists of two steps. In the first part, a CNN is trained to classify strong lens observations as either containing substructure or not. In the second step, the trained CNN is used to quantify the detectability of a single subhalo in every pixel, over a range of masses, in a given mock observation. In this way the method resembles the traditional approach, although without the expensive evidence calculations or forward modelling, these parts having effectively been replaced by the neural network. The mock observations we use in training are produced with realistic source brightness distributions from the Hubble Deep Field \citep{Rafelski2015}. They include elliptical power-law lenses and external shear, as well as a lens light subtraction. 

The mock observations are intended to mimic the Euclid VIS instrument. The excellent angular resolution, seeing, and wide observing area of Euclid makes it particularly attractive for strong lensing studies and so we focus solely on Euclid VIS in this paper. However, the method we propose can readily be adapted to other instruments. A particular focus in this work is whether observations with the resolution of Euclid VIS will themselves be able to constrain the SHMF in a useful way. Subhalo sensitivity is primarily a function of the instrument angular resolution and signal to noise ratio \citep{Despali2022}. Hubble Space Telescope (HST)  observations, at 0.04 arcsec resolution and total S/N $\gtrsim 100$ are sensitive to subhaloes down to a mass of $M_\mathrm{vir} \sim 10^{8}\Msun$ in the best cases, which allows $\Mhm$ to be constrained to roughly the same value \citep{Ritondale2019}. Constraints from other sources, including the abundance of Milky Way satellites, strong lensing flux ratios, gravitational imaging, and the Lyman-$\alpha$ forest, have already ruled out a half mode mass above $\sim10^8\Msun$ \citep[][]{Enzi2021,Nadler2021b,Hsueh2020,Gilman2020}. Even if the sheer number of Euclid VIS images cannot constrain the SHMF below the subhalo mass range of HST images, they will at least provide candidates for follow-up in higher resolution instruments that can. This is another motivation for this work, and for our interest in Euclid in general. If the sensitivity of images can be cheaply estimated a priori, then Euclid strong lenses can be ranked by sensitivity, and only the most promising candidates followed up in higher resolution instruments, e.g. the European Extremely Large Telescope (E-ELT).

The paper is organised as follows. In \cref{sec:mock-observations} we describe the procedure for producing mock strong lens observations. In \cref{sec:method} we detail the machine learning method and the training process. In \cref{sec:results} we present results from simulated Euclid images. In \cref{sec:discussion} we discuss our results and summarise our conclusions. Throughout this paper we assume a Planck 2015 cosmology with $H_0=67.7\,\mathrm{kms}^{-1}\mathrm{Mpc}^{-1}$ and $\Omega_\mathrm{m,0}
=0.302$ \citep{Planck2015}.

\section{Simulated strong lens observations}
\label{sec:mock-observations}

A population of realistic Euclid strong lenses has already been simulated by \citetalias{Collett2015}. As such we follow the procedure therein wherever possible. In this section and the rest of the paper we refer to three types of data: training, testing, and evaluation. Each type uses the same method to produce the data, which we describe in this section, but their underlying model parameters are drawn from different distributions.

Evaluation data is that used to estimate sensitivity statistics for Euclid in our results. Training and testing data are used to train the network, and test its performance during training respectively. The evaluation data is sampled in a realistic way such that distributions of redshifts, Einstein radii, image configuration, signal to noise ratio, etc, are consistent with those expected in nature. These parameters are drawn from a simulation procedure, described below, and compiled in a catalogue. However, when we produce training and testing data we first resample these parameters from uniform distributions wherever possible. The intent is to prevent the neural network from specialising on specific cases that are more common in the physically sampled evaluation dataset. The training and testing data can be thought to cover, as uniformly as possible, all possible strong lenses, whereas the evaluation data are the actual strong lenses that we expect to observe with Euclid.

In training the network, complexity is added to the training data in stages. This section essentially describes the final stage, with previous stages either excluding some part, or using a simplified version of it. The details of these stages are given in \cref{table:training}.

\subsection{Source galaxies}

We begin with a population of sources to be lensed. For this we use the Hubble Deep Field (HDF) catalogue assembled by \citet{Rafelski2015}. To obtain realistic statistics for substructure detectability it is vital to use complex source brightness distributions. This is because the local change in image surface brightness produced by substructure can be absorbed, to an extent, by a change in source surface brightness distribution. A CNN used to detect substructure must take this degeneracy into account, during training, to produce reliable results.

We denoise the sources using the method of \citet{Maturi2017}. The method uses Expectation Maximisation Principal Component Analysis (EMPCA) where images can be reconstructed from principal components. The lowest-order components fit the largest structures in the images, and the highest-order components fit only to the noise. By omitting some number of the largest principal components from the reconstruction, one obtains a denoised version of the original image. We divide the HDF images into six bins according to their angular size given in \citet{Rafelski2015}. We then fit $200$ principal components separately in each bin. We determine the number of principal components to use in the reconstruction individually for each source. To do this, we compute a reduced chi-squared statistic between the reconstruction and the original noisy image, after adding each component. We stop adding principal components when the reduced chi-squared statistic no longer improves. The simplest sources, which are also the most numerous, require $<4$ components for a good fit. The more complicated sources in the sample require up to $40$ principal components. The median number of components used in the entire sample is $11$.

At selection time, the source redshift is resampled from a normal distribution centred on the original redshift in the HDF catalogue with a standard deviation $\sigma_z=0.2$. This helps the network generalise by smoothing out the underlying redshift distribution in the source catalogue, but keeps each source close to its original redshift when imaged. The angular size of the source is adjusted accordingly. The sources are themselves split into a training and testing set, with testing sources also used for the evaluation dataset. None of the sources used in our results were seen by the network during training. The final number of sources used is $4548$ in the training set and $1581$ in the testing and evaluation set.

\subsection{Lens galaxies}

For the lens galaxy we follow the same procedure as \citetalias{Collett2015} and that reference should be consulted for details. We draw a velocity dispersion $\sigmav$ from the elliptical galaxy velocity dispersion function, in the range $[50,400]$kms$^{-1}$, and a lens redshift $\zl$ from the comoving volume function in the range $[0,4]$. The lens is placed in a random angular position $(\lensx,\lensy)$ within the HDF field of view. For all HDF sources in the light cone behind the lens galaxy, the Einstein radius is calculated,

\begin{equation}
	\erad= 4\pi \frac{\sigmav^2}{c^2}\frac{\Dls(\zl,\zs)}{\Dl(\zl)},
\end{equation}
where $c$ is the speed of light and $\Dls$ and $\Dl$ are the angular diameter distances between the lens and the source, and the observer and the lens respectively. The source position $\beta$ is found relative to the centre of the lens galaxy and sources with $\beta<\erad$ are accepted.

The lens galaxy mass profile is a singular power-law ellipsoid. The dimensionless projected surface mass density, or convergence, is given by
\begin{equation}
	\kappa(\etheta)=\frac{2-t}{2}\left(\frac{b}{\etheta}\right)^{1-t},
\end{equation}
where $t$ is the slope of the mass profile and $b$ is the lensing strength, where $b=\erad\sqrt{q}$ and $q$ is the axis ratio of the elliptical mass distribution \citep{Tessore2015}. The axis ratio $q$ also defines the elliptical radius $\etheta^2=\theta_x^2q^2 + \theta_y^2$. The axis ratio is drawn from the axis ratio function for elliptical galaxies, as in \citetalias{Collett2015}, and $b$ can then be set from the already drawn velocity dispersion and Einstein radius. The slope of the mass profile is drawn from a normal distribution centred on $t=1$ (an isothermal slope) with standard deviation $\sigma_t=0.1$, approximately the distribution of slopes in lens galaxies \citep{Bolton2008}. We use a S\'ersic profile for the lens galaxy surface brightness distribution with effective radius $\reff$ and magnitude $\Ml$ drawn from the fundamental plane relation for elliptical galaxies. The S\'ersic index is drawn from a scaling relation with the magnitude. Finally, to simulate the lensing effect of objects in the proximity of the main lens, we add external shear with a random angle and a random strength between $0.0$ and $0.1$. This is typical of the shear strength found in strong lens observations \citep{Bolton2008}.

\begin{figure*}
	\includegraphics[width=1.0\textwidth]{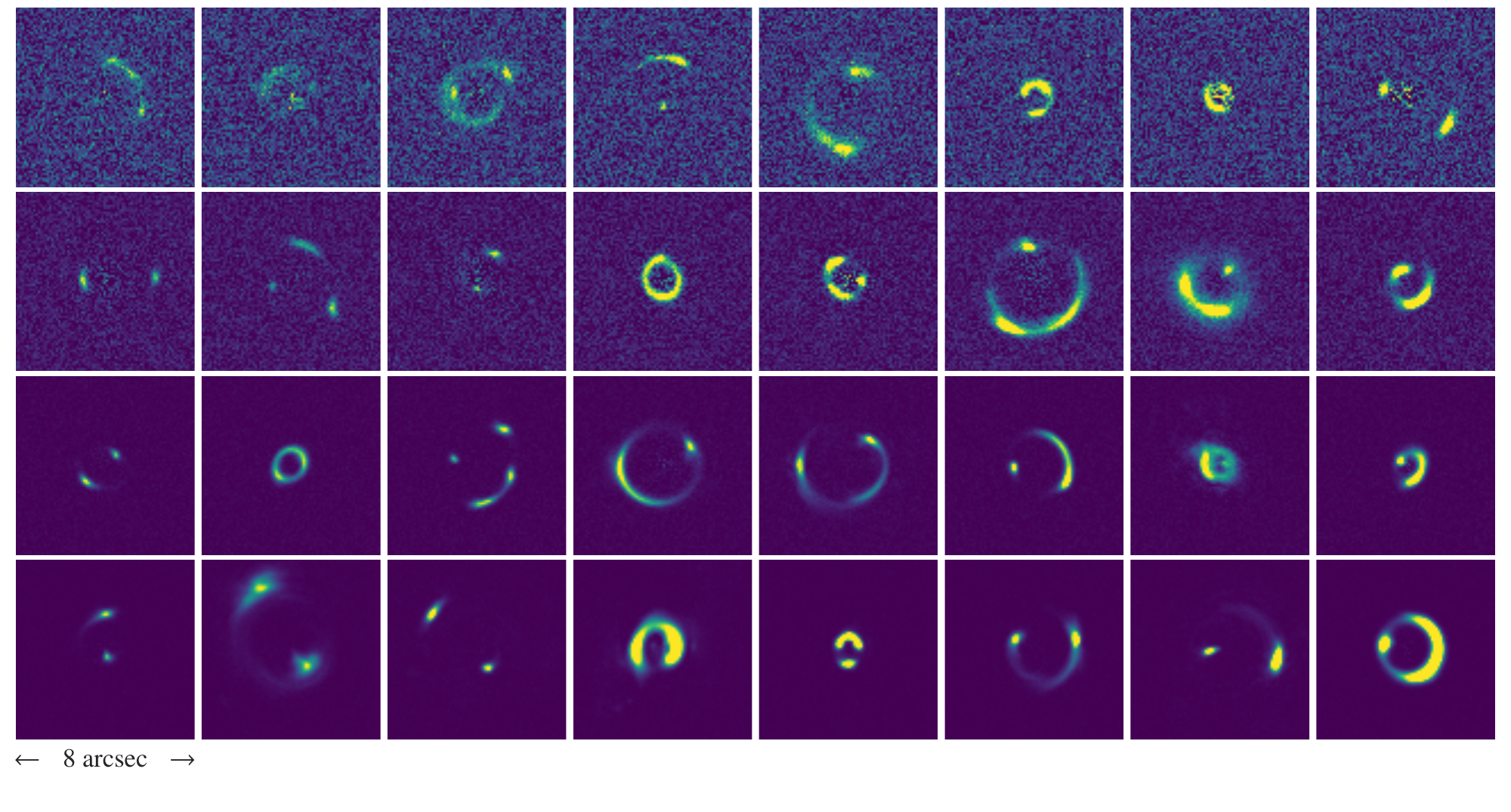}
	\caption{\label{fig:training-data} A representative sample of the simulated strong lens observations used in the final stage of training (see \cref{table:training}). These observations contain all the ingredients described in \cref{sec:mock-observations}. The lens light has been subtracted. Systems are ordered in the figure by their maximum brightness. To adequately show detail at different signal to noise ratios, each row is normalised with a different colour scale. As training observations, these data are not intended to be realistically sampled, especially in signal to noise ratio. The upper two rows most resemble the more realistic Euclid VIS data that we use in \cref{sec:results}.}
\end{figure*}

\subsection{Subhaloes} 

In the training and testing data we add, with equal probability, either: zero subhaloes or; one to four subhaloes. The subhaloes are placed uniformly in a 6 arcsec $\times$ 6 arcsec square centred on the lens galaxy. The largest lens used has $\erad<3$ arcsec, so the subhalo area always encloses $\erad$. For the subhaloes we use an NFW density profile,
\begin{equation}
	\label{eq:nfw-3d-density}
	\rho(r)=\frac{\rhos}{(r/\rs)(1+r/\rs)^2},
\end{equation}
where $r$ is the radius in three dimensions, $\rs$ is the scale radius and $\rhos$ is the normalisation, itself given by
\begin{equation}
	\rhos=\frac{\vmax^2 C(1+C)^2}{4\pi G\rmax^2},
\end{equation}
where $\vmax$ is the maximum circular velocity of particles in the subhalo and $C$ is a constant that relates $\rs$ to $\rmax$, the radius at which $\vmax$ occurs,
\begin{equation}
	\label{eq:rmax-rs-relation}
	C=\frac{\rmax}{\rs}\approx 2.163.
\end{equation}
Typically, the mass of the subhalo $\Msub$ is taken to be the mass enclosed within the subhalo virial radius, $r_\mathrm{vir}$ prior to infall, which is labelled $M_\mathrm{vir}$. In this work we instead use $\Msub=\Mmax$, the mass enclosed by $\rmax$. This more closely approximates the masses of subhaloes in simulations.

In training and testing the subhalo mass is drawn from a log-uniform distribution with limits $10^{8.6}\Msun\leq\Mmax\leq10^{11}\Msun$. The concentration of the halo is typically set by the concentration mass relation given by \citet{Duffy2008}. This relation for subhaloes is an extrapolation from that for larger haloes and may not be accurate at smaller masses, underestimating the concentration for subhaloes.

Instead, we set the concentration by applying a $\vmax$-$\rmax$ relation obtained using the ShinUchuu simulation ~\citep{Ishiyama2020,Moline2022},
\begin{equation}
	\label{eq:rmax-vmax-relation}
	\rmax = A\left(\frac{\vmax}{10\,\mathrm{kms}^{-1}}\right)^B,
\end{equation}
where $A=0.344\,\mathrm{kpc}$ and $B=1.607$\footnote{This version of the relation does not include a redshift dependence, although more recent versions do, see \citet{Moline2022}.}. ShinUchuu is a higher resolution simulation with $6400^3$ particles covering a 140 $h^{-1} \rm Mpc$ side length box, with resulting mass resolution of $8.97 \times 10^5 \, h^{-1}\, \mathrm{M_{\odot}}$. Halo catalogues are available for public download\footnote{\url{http://www.skiesanduniverses.org}} and we use the results for $\vmax$ and $\rmax$ . We grouped subhaloes in bins of $\vmax$ of equal size and obtained the mean of $\rmax$.  Based on these results, we propose the $\vmax$-$\rmax$  relation of \cref{eq:rmax-vmax-relation}. The parametrization works well for subhaloes with $\vmax$ between $38$~km s$^{-1}$ and $300$~km~s$^{-1}$ in host haloes with masses between $\sim 9.8\times 10^{12} h^{-1} \Msun$ and $1.2\times 10^{13}~h^{-1} \Msun$ (its accuracy being better than 5 per cent at all $\vmax$ values within this range).

In training and testing, we use the relation to set limits on a log-uniform distribution with $1.5\,\mathrm{kpc}<\rmax<28.0\,\mathrm{kpc}$, from which we draw a random $\rmax$. In evaluation, we convert the chosen $\Mmax$ to $\vmax$ and take $\rmax$ directly from \cref{eq:rmax-vmax-relation} for the given $\vmax$. The NFW scale radius is then given in either case by \cref{eq:rmax-rs-relation}. It is important to note that this relation produces subhaloes which are more concentrated than those used in previous gravitational imaging studies \citep[e.g.][]{Despali2018}. To make this clear, subhalo characteristics in our mass range are printed in \cref{table:concentrations}.

\begin{table}
	\begin{tabular}{ p{1cm} p{1cm} p{1cm} p{1cm} p{0.9cm} p{0.9cm} }
		\hline
		$M_\mathrm{vir}$&$\Mmax$& $\vmax$& $\rmax$& $c_\mathrm{vir}$ & $c_\mathrm{vir}*$\\
		$\left[\Msun\right]$ &$\left[\Msun\right]$ & [kms$^{-1}$] &[kpc] &\\
		\hline
			$10^{8.0}$ & $10^{6.2}$ & $6.3$ & $0.16$ & $13.5$ & $60.2$\\
			$10^{9.0}$ & $10^{7.4}$ & $13.6$ & $0.56$ & $11.2$ & $42.5$\\
			$10^{10.0}$ & $10^{8.6}$ & $29.2$ & $1.92$ & $9.3$ & $29.9$\\
			$10^{11.0}$ & $10^{9.8}$ & $62.6$ & $6.56$ & $7.7$ & $20.9$\\
			$10^{12.0}$ & $10^{11.0}$ & $134.0$ & $22.27$ & $6.4$ & $14.6$\\
		\hline
	\end{tabular}
\caption{\label{table:concentrations} Characteristics of the subhaloes used in this work, using an example redshift $z=0.5$. The first column is a given virial mass, and the next three columns give the equivalent $\Mmax$, $\vmax$ and $\rmax$ for a subhalo using our relation in \cref{eq:rmax-vmax-relation}. The final two columns are: the concentration $c_\mathrm{vir}$ that a subhalo with the given $M_\mathrm{vir}$ would have according to \citet{Duffy2008}, and, the concentration $c_\mathrm{vir}*$ that the equivalent subhalo has using \cref{eq:rmax-vmax-relation}, and assuming an NFW profile.}
\end{table}

\subsection{Instrument characteristics}
\label{sec:instrument}

Using the described sampling procedure, we ran simulations to collect a large number of lens and source parameters. From this catalogue we then build the mock observations for training, testing and evaluation. The source surface brightness distribution is ray-traced through the mass model onto a grid with pixel size $0.1$ arcsec and a field of view of $10$ arcsec. To ensure an accurate source reconstruction in the image plane, we subsample each pixel with $10\times10$ subpixels and use the mean over these subpixels. We add the lens light and convolve with a Gaussian PSF with FWHM $=0.16$ arcsec. To this we add a uniform sky brightness of $M_\mathrm{VIS}=22.2$. We then compute the total expected counts for an observation with a zero-point of $M_\mathrm{VIS}=25.2$ and an exposure time of $3\times 565=1695$ s. The Euclid Wide Survey uses four exposures of $565$ s each, although due to gaps in the detector array and other technical considerations, the full survey area is only covered by three exposures \citep{Scaramella2022}. VIS technical details are taken from \citet{Vavrek2016} and \citet{Cropper2018}. We use the expected counts in each pixel as the mean of a Poisson distribution from which we draw the actual total counts. Finally, we subtract the original S\'ersic profile describing the lens light from this noisy image, leaving a Poisson-limited lens light subtraction. An example set of training observations are shown in \cref{fig:training-data}.

\section{Method}
\label{sec:method}

Our method comprises a number of steps. First, a neural network is trained on realistic mock observations that contain either zero or between one and four subhaloes, randomly placed in the image. The training is performed in stages where the complexity of the data is gradually increased and the model retrained. In the training stage, the observation parameters (Einstein radius, signal to noise ratio etc) are drawn uniformly from a wide range. Second, the trained model is used to predict the sensitivity in a second set of observations. In this stage, the observation parameters are drawn from realistic distributions, intended to match the expected population of Euclid strong lenses. For a given system in this population, we produce realisations of the same observation with a single subhalo, iterating over all subhalo positions and masses. For every position and mass, the trained network gives the probability that a subhalo is in the image. By defining a probability threshold at which a detection would be acceptable, we obtain the minimum detectable subhalo mass in each pixel. We repeat this process for a large number of mock Euclid observations to obtain our results.

\subsection{Machine learning}

We use the ResNet-50 architecture, commonly used in image classification tasks \citep{He2015}. ResNet is a residual convolutional neural network that utilises skip connections between network layers. These skip connections are designed to overcome the `vanishing gradient' problem, and allow for the training of very deep networks. The architecture we use is unmodified from the original 50 layer implementation so we defer to the previous reference for details.

For a given image $d$, the neural network returns the probability $\prob{C=i}{d}$, that the image belongs to class $i$. In our case there are only two classes, $C=0$ for an image with no substructure, and $C=1$ for an image with one or more substructures. We train using the Adam optimizer and minimise the cross-entropy between the network's predictions and the truth. We use a batch size of $1024$. At training time, images are rescaled to the range $[0,1]$, then randomly rotated, flipped and cropped to a size of $8$ arcsec, producing $80\times80$ sized images from the original $100\times 100$. The random crop ensures that the lens is not at the centre of the image, and improves training performance, without discarding any useful information given that all lenses have $\erad<3$ arcsec.

Training takes place in stages, with slight increases in the difficulty of the task at each stage. The stages are detailed in \cref{table:training} and each stage's data set consists of $2\times10^6$ images. To achieve changes in the range of total signal to noise ratios, source magnitudes were changed from an initially constant $M_\mathrm{VIS}=20$ in stage one, to a uniformly sampled range $20<M_\mathrm{VIS}<26$ in stage three onwards. The lower limit of the subhalo mass range was moved down to $10^{8.6}\Msun$ in stage five from $10^{9}\Msun$ in stages one to four once it was found that a small number of lenses had sensitivity below this limit. External shear is added in stage four. At the start of each new stage we conduct a parameter search to find the optimum learning rate and the network starts with the converged parameters from the previous stage. As training progresses, the learning rate is multiplied by a decay factor of $10^{-0.25}$ whenever the testing loss has not decreased for ten epochs. The network is assumed to have converged when three decreases in learning rate do not improve the test loss.

\begin{table}
	\begin{tabular}{ l l l l l l l l}
		\hline
		   	& Min. S/N   & Min. $\Mmax/\Msun$ & E.S. 		& $N_\mathrm{sub}$			& Loss   & Acc.   \\
		\hline
		1  	& $10^2$     & $10^{11}$          & $0.0$       	& 0 or 1					& 0.307 & 0.867	\\
 		2  	& $10^2$     & $10^9$             & $0.0$       	& 0 or 1			    	& 0.575 & 0.673	\\
		3  	& $20$       & $10^9$             & $0.0$    		& 0 or 1				    & 0.610 & 0.638	\\
		4 	& $20$       & $10^9$             & $0.1$       	& 0 or 1				    & 0.641 & 0.599	\\
		5 	& $20$       & $10^{8.6}$         & $0.1$       	& 0 or 1				    & 0.678 & 0.538	\\
		6 	& $20$       & $10^{8.6}$         & $0.1$       	& 0 or 1-4					& 0.658 & 0.569 \\
		\hline
	\end{tabular}
	\caption{\label{table:training} Changes in the training data during training. The columns are: minimum signal to noise ratio, minimum subhalo mass, maximum external shear strength, number of subhaloes, final testing loss, and, final testing accuracy. The maximum signal to noise ratio in all stages is $10^3$ and the maximum subhalo mass in all stages is $10^{11}\Msun$.}
\end{table}

\subsection{Model performance}
\label{sec:model-performance}
The performance of the model on the testing data at the end of each training stage is given in the final columns of \cref{table:training}. The performance gradually degrades as the problem gets more difficult, aside from in the final stage where the addition of multiple subhaloes makes the problem slightly easier. After the final stage the performance is only slightly better than a random binary classifier, which would have a loss of 0.693 and an accuracy of 0.5. However, this is to be expected considering the difficulty of the problem. Most of the subhaloes in the training and testing data are simply undetectable with any method for data of this quality. Primarily this is because they are too small, or too far from the lensed images to have a detectable effect on the local deflection angle. Even if a massive halo is in the right position, an unfavourable signal to noise ratio, source structure, lens or source redshift, or subhalo concentration could all prevent it from being detected.

It is more instructive to evaluate the performance in specific situations where we expect positive classifications to be possible. We test the network's response to specific subhalo positions $(x_\mathrm{sub},y_\mathrm{sub})$ and masses $\Msub$ by creating many realisations of the same system where the subhalo position and mass change, but everything else stays fixed. The macro properties of the observation: lens light and mass model, source model, sky noise realisation, and external shear, are kept the same. Passing each realisation through the network gives the probabilities $\prob{C=0}{d}$ for no substructure and $\prob{C=1}{d}$ for any substructure, where $d$ is the image for the realisation with that subhalo position and mass. The detection significance $\detsig$ for a subhalo of a given position and mass is then
\begin{equation}
	\label{eq:significance}
	\detsig(x_\mathrm{sub},y_\mathrm{sub},\Msub)=\sqrt{2}\erf^{-1}\left[\prob{C=1}{d}\right],
\end{equation}
where $\erf^{-1}$ is the inverse error function.

\begin{figure}
	\includegraphics[width=1.0\columnwidth]{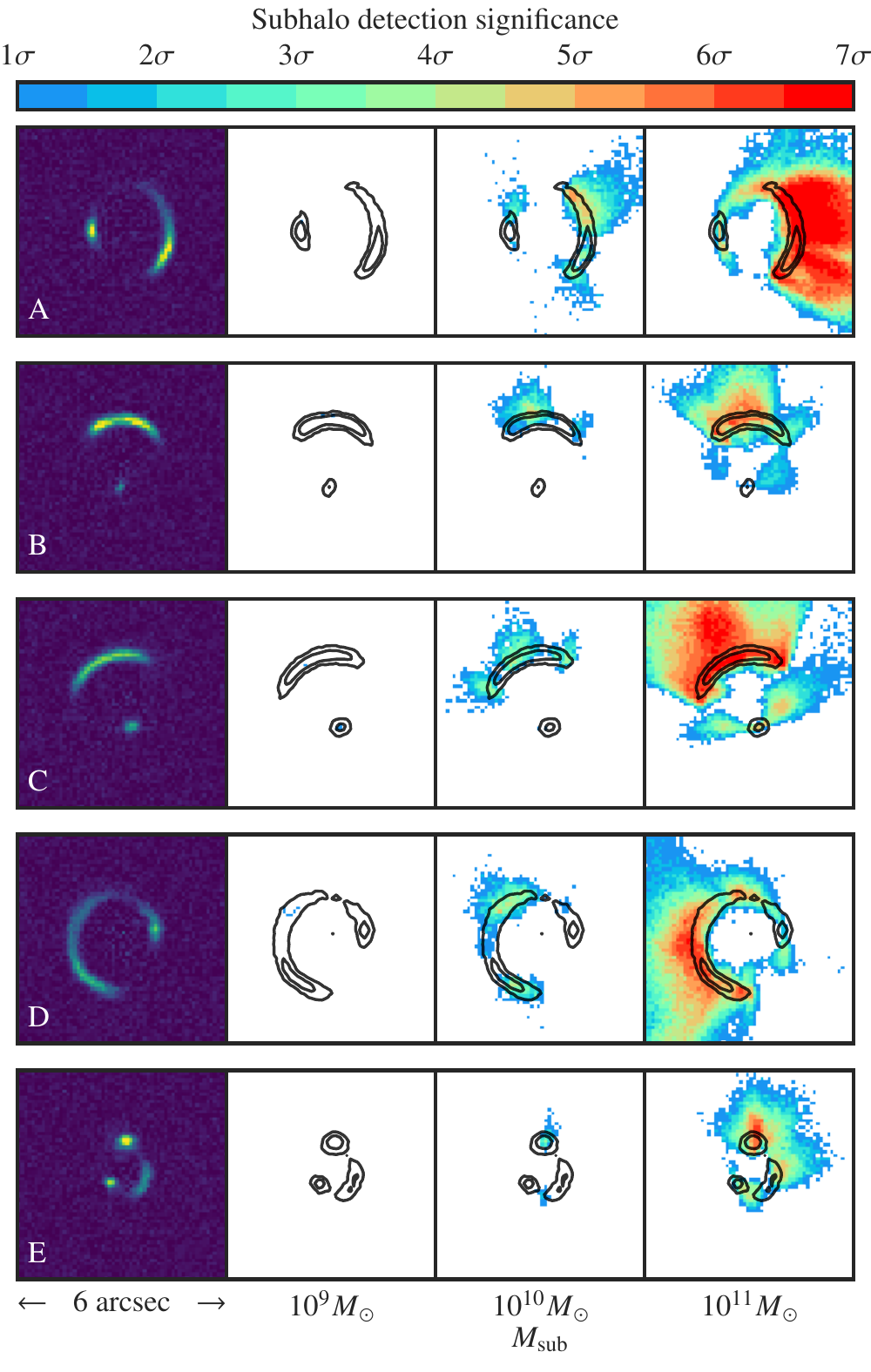}
	\caption{\label{fig:accuracy-maps} The detection significance returned by the trained network as a function of subhalo mass and position in five different systems. Observations are from the evaluation set and share the same colour scale. White pixels on the significance maps indicate a detection significance below $1\sigma$. Black contours outline the lensed images. Subhalo mass is labelled along the bottom.}
\end{figure}

\begin{figure}
	\includegraphics[width=1.0\columnwidth]{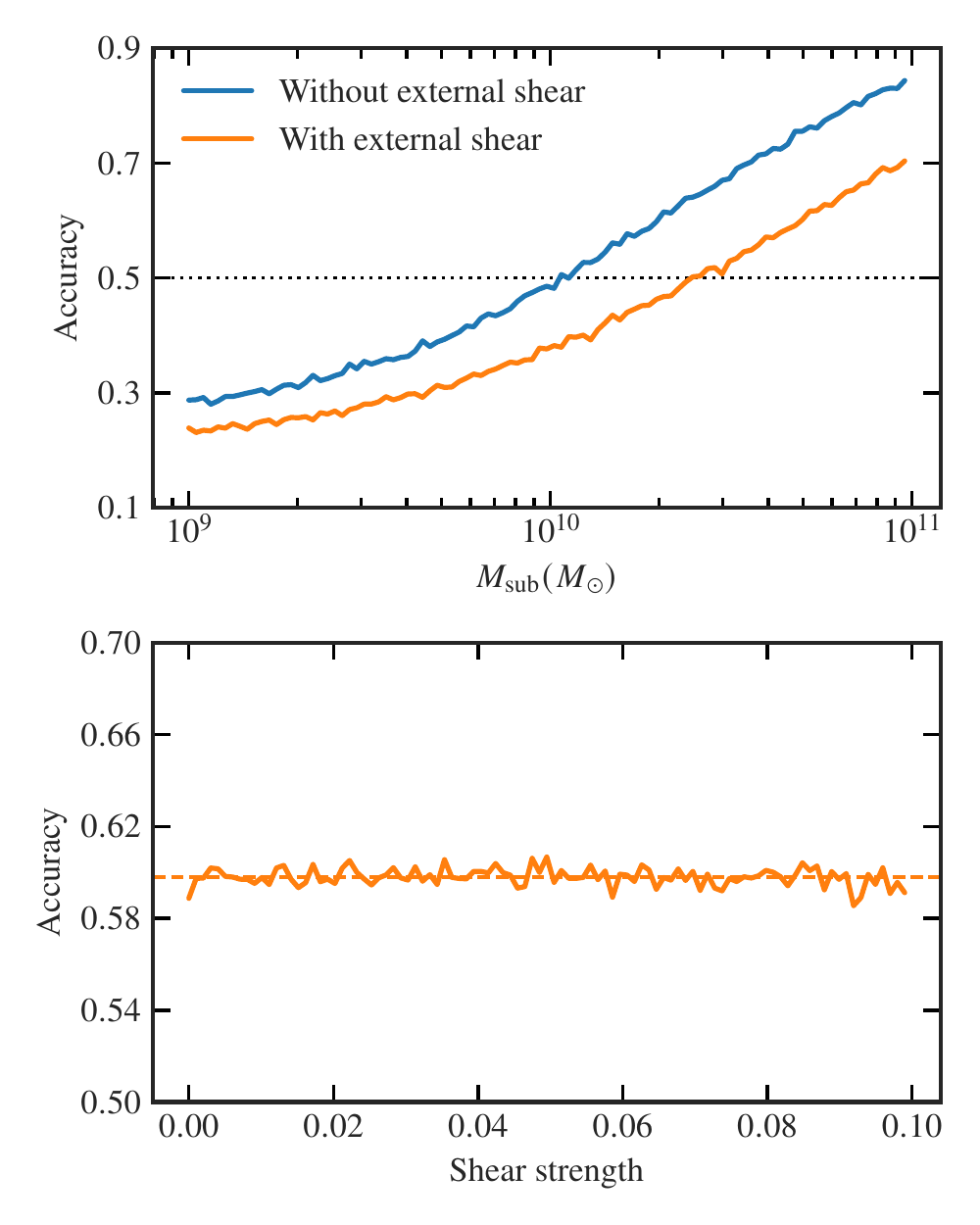}
	\caption{\label{fig:shear-strength} The effect of adding external shear to the training data on the model performance. The upper frame shows the testing accuracy for positive classifications only, binned by subhalo mass. Stage 3 data contains no external shear, stage 4 contains external shear with a random strength varying uniformly from $0.0\leq|\gamma|\leq0.1$. The lower frame shows the accuracy for all (positive and negative) data in stage 4 binned by shear strength. The dashed line is the accuracy over the entire dataset.}
\end{figure}

In \cref{fig:accuracy-maps} we plot maps of this significance for five different example systems. We iterate the subhalo position across all image pixels in the central 6 arcsec $\times$ 6 arcsec area and sample three masses of $\Mmax=\{10^{9},10^{10},10^{11}\}\Msun$. The figure shows some general behaviour common to all systems. Areas away from the lensed images do not produce detections, except at very high masses, and detections close to or on the lensed images are easier. This is also the case in traditional modelling techniques \citep{Minor2021,Nadler2021b}. Despite the statistics in \cref{table:training}, we see that the model performs well in situations where a detection should be physically possible. Systems B, C, and D have a very small number of pixels where a $10^9\Msun$ subhalo can be detected with low significance, but the performance improves rapidly with subhalo mass. At $10^{10}\Msun$, all systems except E have detections at $5\sigma$, and significant areas with detections above $3\sigma$.

Comparing the performance of the method with the traditional forward modelling approach of e.g. \citet{Vegetti2009} at this stage is difficult. This and previous gravitational imaging studies use different definitions of subhalo mass, different subhalo concentration relations, different definitions of sensitivity (e.g. we account for the presence of multiple subhaloes), and have fundamentally different priors. A direct comparison with the traditional method then requires modifications to that method, and that many computationally expensive sensitivity maps are reproduced with all the required similarities. As such, we defer this comparison to a future paper.

\begin{figure*}
	\includegraphics[width=1.0\textwidth]{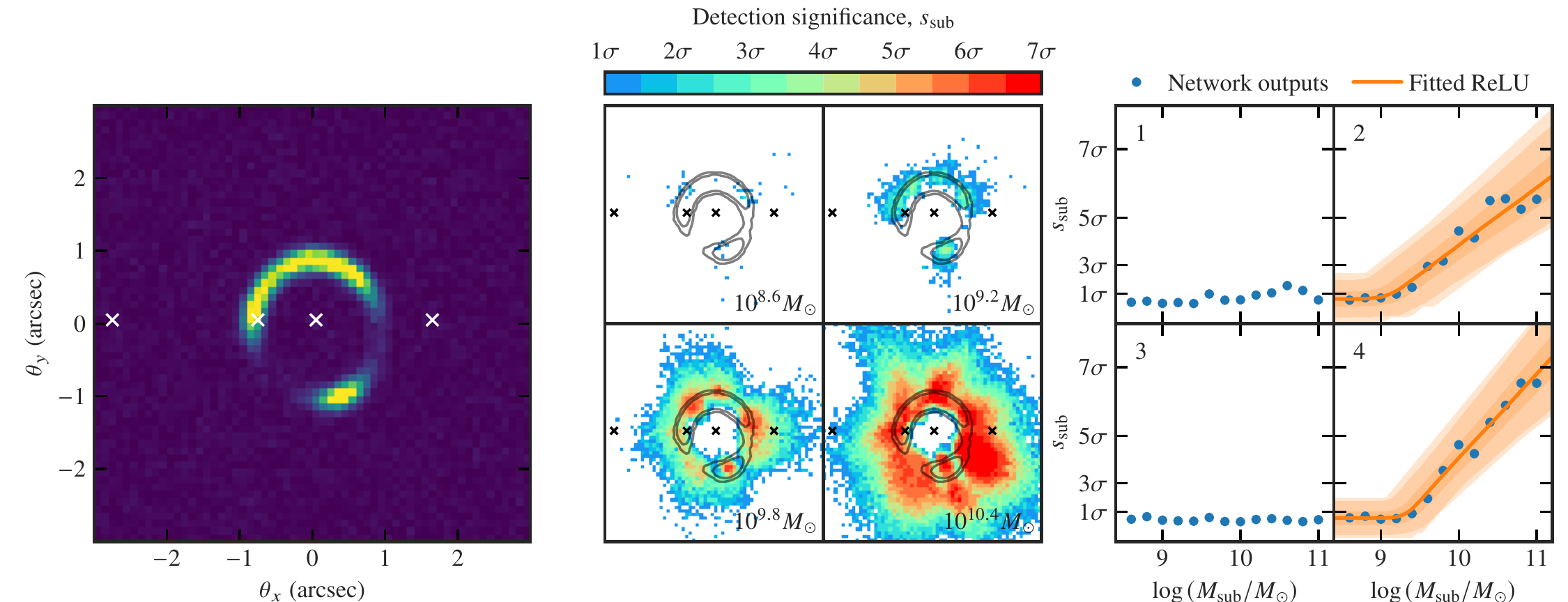}
	\caption{\label{fig:sensitivity-estimation} The procedure for estimating the sensitivity in an example system, shown in the left-hand frame. In the central frame is the detection significance $\detsig$ at four different subhalo masses in every pixel. Contours showing the lensed image positions are over-plotted. The right-hand frame shows the significance $\detsig$ at all $13$ mass steps in four example pixels, labelled one to four with white (black) crosses in the left-hand (middle) frame. For simplicity, the pixels all sit on a straight line through the centre of the lens. The orange curve shows the best fit ReLU function from \cref{eq:relu}. The orange shaded area shows the $64\%$, $95\%$ and $99\%$ confidence regions for the fit, obtained by sampling ReLU realisations using the uncertainties on the fitted parameters.}
\end{figure*}

\begin{figure}
	\includegraphics[width=1.0\columnwidth]{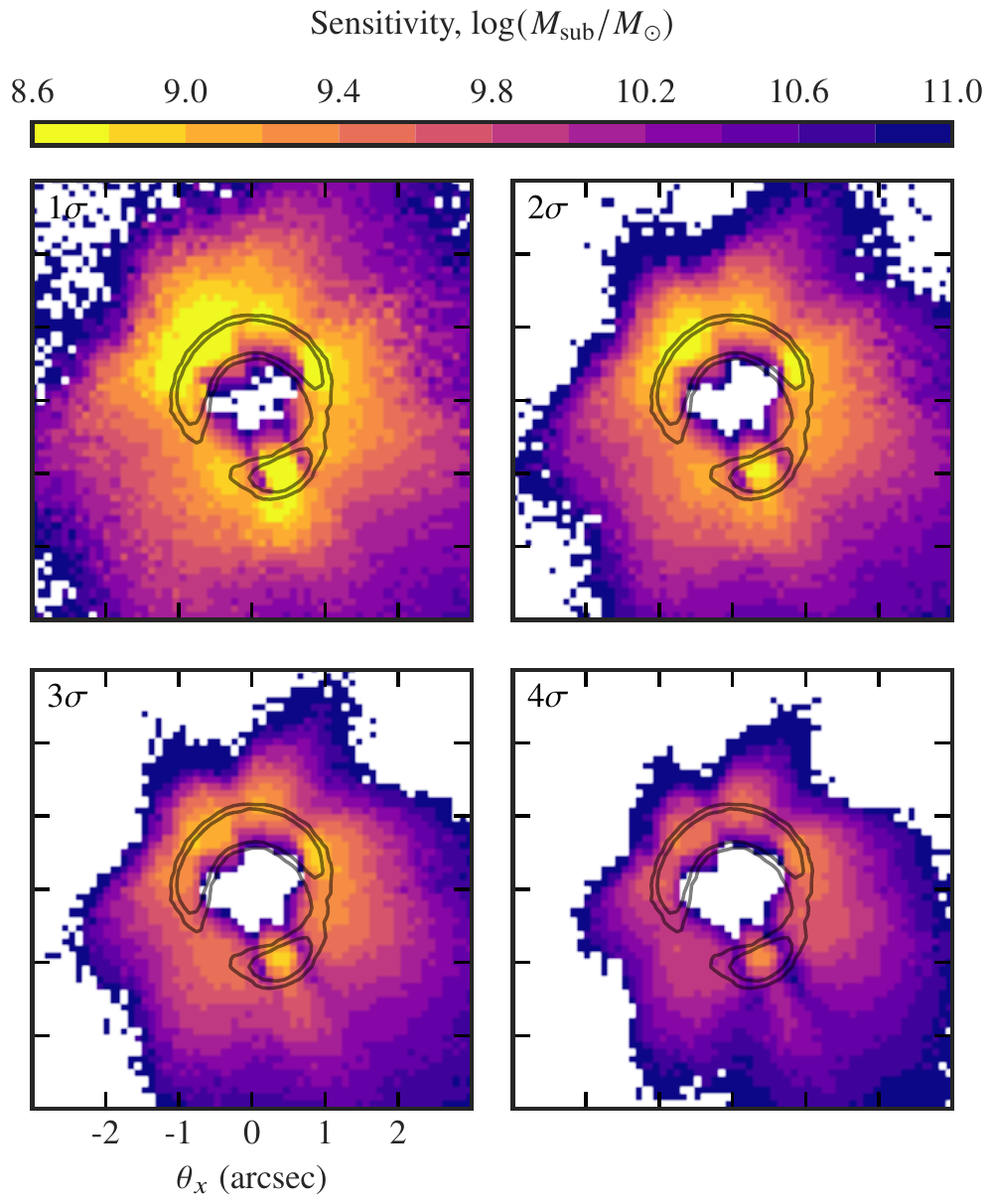}
	\caption{\label{fig:sensitivity-example} Sensitivity maps at four different detection thresholds for the example system in \cref{fig:sensitivity-estimation}. Contours show the position of the lensed images. White pixels in the map indicate those where no sensitivity was found, either because the detection threshold was not reached by any mass in that pixel, or because the minimum detectable mass was outside the sampled range.}
\end{figure}

In all systems in \cref{fig:accuracy-maps} we see that subhalo detectability diminishes towards the centre of the lens, i.e., inside the lensed images. In fact, at the very centre, it is impossible to detect a subhalo of any mass. This is because a subhalo in this position has the same effect on the lensed images as adding a small amount of mass to the lens galaxy mass model, parametrically equivalent to increasing the Einstein radius. This degeneracy is also present in traditional modelling techniques. The network has successfully learned this degeneracy from the training data, although it was not explicitly required to do so.

A similar phenomenon is observed when we add external shear to the training data. \Cref{fig:shear-strength} shows the impact on model performance when we introduce external shear. A subhalo close to the lensed images will produce a local shearing effect. The external shear added in stage four to mimic the effect of larger nearby objects can replicate this local shear to an extent, depending on configuration.

The network has learned this degeneracy, shown by the drop in accuracy across all masses in \cref{fig:shear-strength}, as all of its predictions are now less confident. The size of this drop is slightly larger at larger masses, where the shear produced by a subhalo is less localised, and so, more easily replicated by a global external shear. Importantly, the accuracy in stage four does not depend on the strength of shear in a given image as we see in the lower panel of \cref{fig:shear-strength}. If the network was confusing systems with a strong external shear for those with a subhalo (or vice versa) we would see a drop in accuracy as shear strength increases, rather than the constant accuracy plotted here.

\begin{figure*}
	\includegraphics[width=1.0\textwidth]{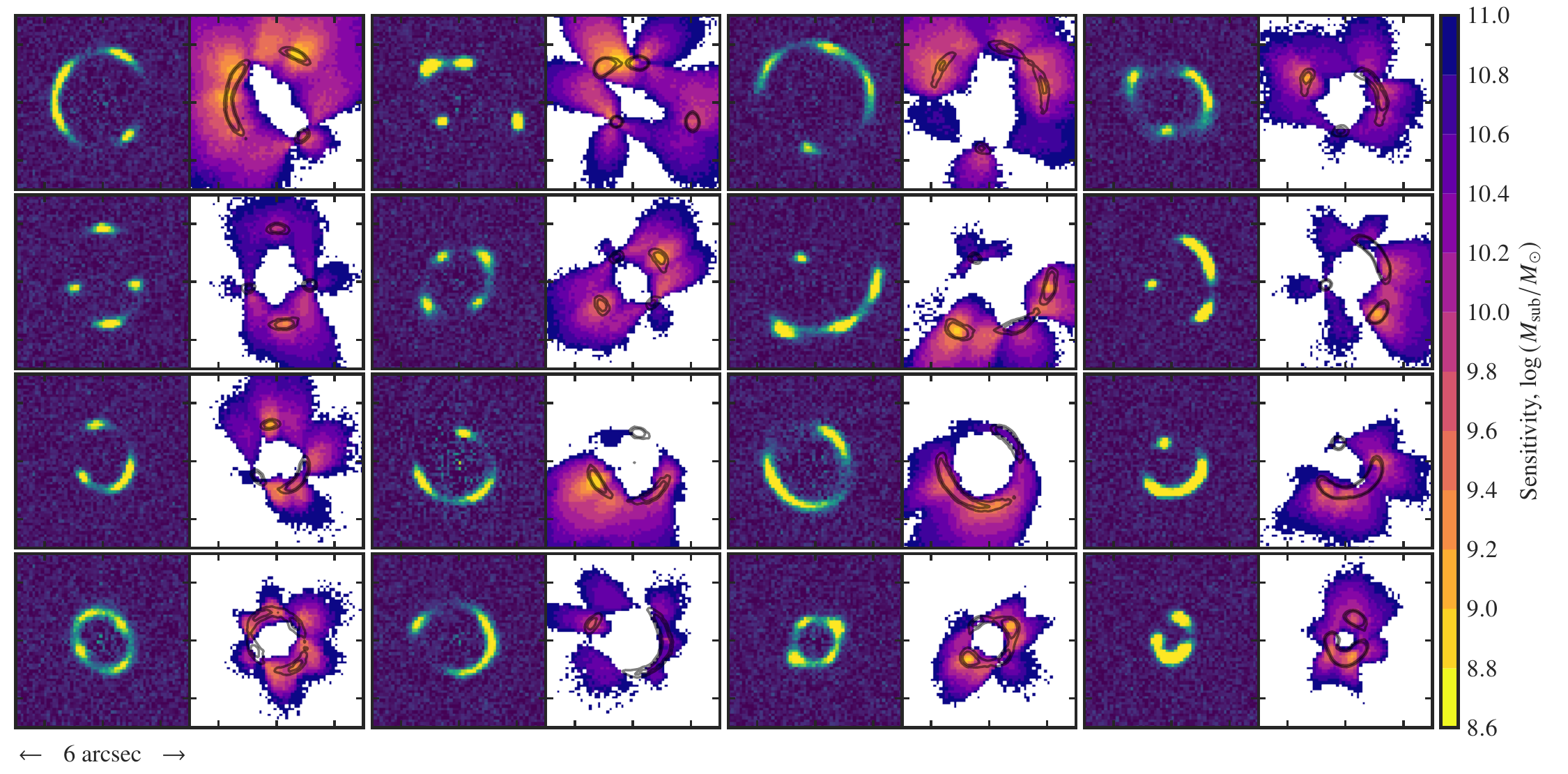}
	\caption{\label{fig:sensitivity-maps} Mock observations and their $3\sigma$ sensitivity maps chosen from the most sensitive $10$ per cent of systems. All observations share the same colour scale. Contours on the sensitivity maps indicate the location of the lensed images. White pixels are those with no sensitivity in the tested range, i.e., no subhalo of any mass was detected at $3\sigma$.}
\end{figure*}

\subsection{Sensitivity estimation}
\label{sec:sensitivity-estimation}
We can now use our trained model to produce sensitivity maps for the simulated observations in our evaluation data. The process is similar to that used to produce the maps in \cref{fig:accuracy-maps}.  For a particular system in our simulated catalogue, we produce mock observations of that system with a subhalo in each pixel, over a range of masses of interest. At each iteration the subhalo is placed in the centre of the pixel, with a concentration $\rmax$ given by \cref{eq:rmax-vmax-relation}. The system is ray-traced as in \cref{sec:instrument} and the resulting image is evaluated by the network, giving the probability of a subhalo existing in the image, for that subhalo position and mass. When the subhalo position and mass are iterated over, the sky noise realisation, lens mass model, external shear, and lens light distribution are kept the same. As before we use every pixel in the central $6$ arcsec $\times$ $6$ arcsec area, but expand the mass range to $10^{8.6}\leq\Mmax/\Msun\leq10^{11}$, with 13 mass steps uniformly distributed in log-space. This means we are required to ray-trace and evaluate $60\times 60\times 13=46,800$ realisations of each system. This takes $\sim30$ minutes per system using one A100 GPU.

\Cref{fig:sensitivity-estimation,fig:sensitivity-example} illustrate the process for computing a sensitivity map from these realisations and probabilities for one example system, shown in the left-hand frame of \cref{fig:sensitivity-estimation}. In each pixel we are required to find the smallest subhalo which can be detected at a given significance. To do this we fit a rectified linear unit (ReLU) function to the log-odds of a detection as a function of mass, in every pixel. The log-odds increases linearly with subhalo mass and so is more useful here than the significance. The log-odds $R$ is defined as
\begin{equation}
	\label{eq:log-odds}
	R=\log\left[\frac{\prob{C=1}{d}}{\prob{C=0}{d}}\right],
\end{equation}
and can be converted to significance using \cref{eq:significance} and the fact that $\prob{C=1}{d}=1-\prob{C=0}{d}$. This is because the probabilities are the output of a softmax function in the final layer of the neural network. The ReLU function we use is
\begin{equation}
	\label{eq:relu}
	R(\Mmax)=\max\left[R_0,a \log\left(\Mmax-M_0\right)+R_0\right],
\end{equation}
where $R_0$, $M_0$ and $a$ are constants found in the fitting process. The uncertainty on $R(\Mmax)$ can be found using the uncertainties on the fitted constants and error propagation of \cref{eq:relu}.

The right-hand frame of \cref{fig:sensitivity-estimation} illustrates the fitting process in four example pixels. For each pixel we first check that the required detection threshold $\detsig$ has been met for any mass. If not, a fit is not performed. The threshold, $3\sigma$ in this case, is not reached for any mass in pixels (1) and (3), because they are far away from the lensed images, and in the centre of the lens respectively (see \cref{sec:model-performance}). They are labelled as having no sensitivity in the sampled mass range.

Pixels (2) and (4) show typical behaviour for sensitive pixels. At low masses, the network returns a probability of substructure close to $50$ per cent. It is important to note that the network will never assign a strong probability to class $C=0$ (no substructure) because the training data contained many examples labelled $C=1$ where the subhalo was undetectable. In the majority of cases, the two classes are indistinguishable. Strong probabilities are then only ever assigned to detections, not non-detections. After a large enough subhalo mass is reached, the network probability increases linearly with log-mass. By fitting \cref{eq:relu} in every sensitive pixel and inverting for $\Mmax$ we can construct a sensitivity map for a given $\detsig$. The final sensitivity maps for our example system are plotted in \cref{fig:sensitivity-example}.

\section{Results}
\label{sec:results}

We generate $16$,$000$ simulated Euclid-like strong lens observations following the procedure in \cref{sec:mock-observations} and compute sensitivity maps with the method in \cref{sec:method}. We impose the same cuts as \citetalias{Collett2015}, namely, lenses must have an Einstein radius $\erad>0.5$ arcsec and a total signal to noise ratio $\mathrm{S/N}>20$. The sample size is approximately one tenth the size of the predicted Euclid strong lens sample \citepalias{Collett2015}. We perform two checks on the completed sensitivity maps before computing the rest of the results. First, we remove any system where a single pixel has a poor fit to \cref{eq:relu}, by checking the uncertainty on the sensitivity. This is a conservative step because only $0.017$ per cent of pixels fail in this way, but $2.1$ per cent of systems have a failed pixel. Second, we test for false positive detections by checking the fitted value of the $R_0$ parameter in \cref{eq:relu}. If $R_0$ is greater than the threshold, calculated with \cref{eq:log-odds}, in every pixel in a system, then it is also removed. A further $0.27$ per cent of systems are removed in this way. The final sample size is then 15,618 lenses. A small number of example sensitivity maps from the sample are plotted in \cref{fig:sensitivity-maps}.

\subsection{Sensitivity statistics}

\begin{figure}
	\includegraphics[width=1.0\columnwidth]{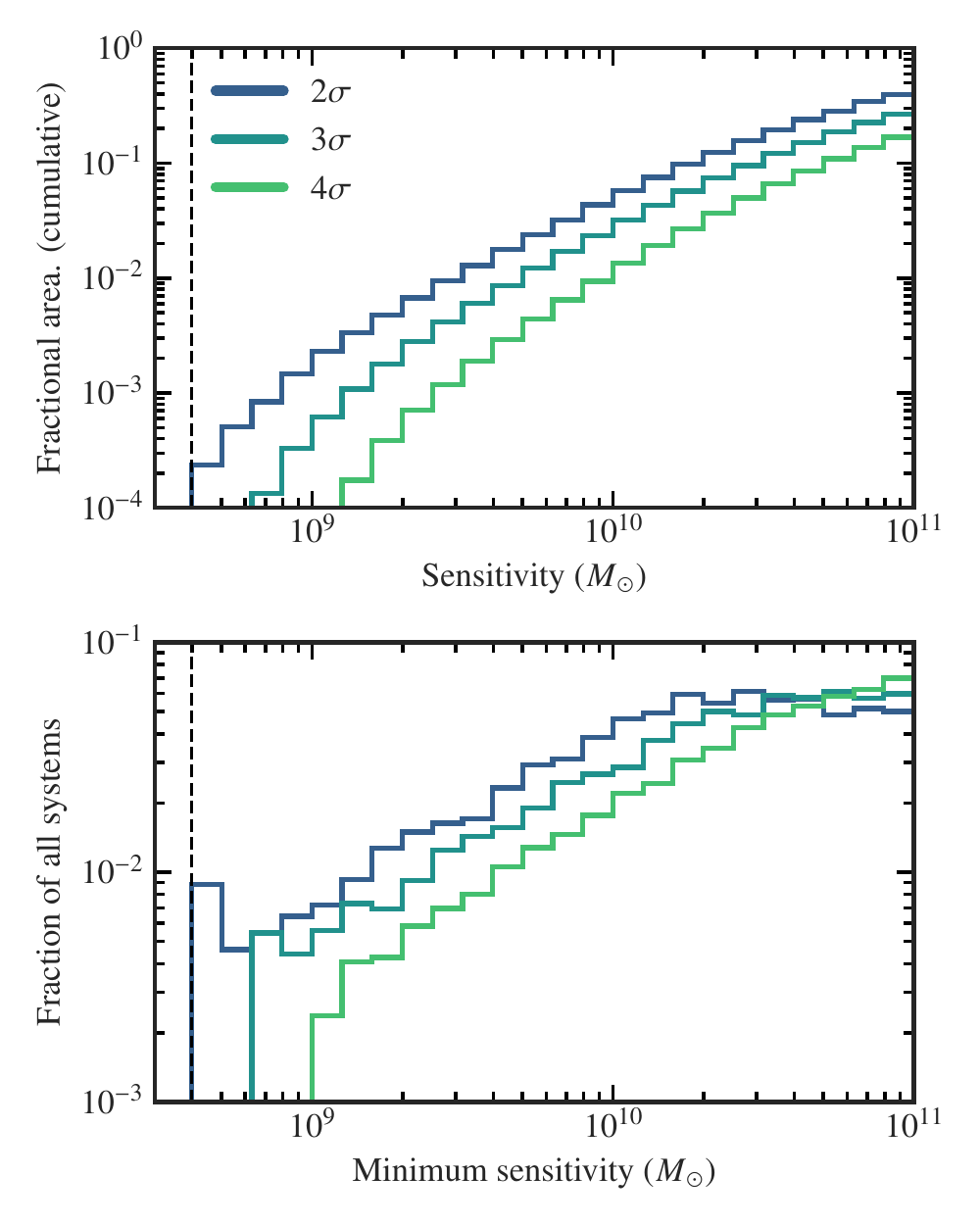}
	\caption{\label{fig:sensitivity-statistics} Statistics for subhalo mass sensitivity at three detection thresholds (coloured). Upper frame: The cumulative distribution of fractional area in all strong lens systems as a function of subhalo mass sensitivity. Lower frame: the distribution of the minimum sensitivity in each system, as a fraction of all systems. At $3\sigma$, the most sensitive pixel has $\Mmax=10^{8.8\pm0.2}\Msun$. In both frames the dashed vertical line indicates the lower limit on subhalo mass used in training.}
\end{figure}

\begin{figure}
	\includegraphics[width=1.0\columnwidth]{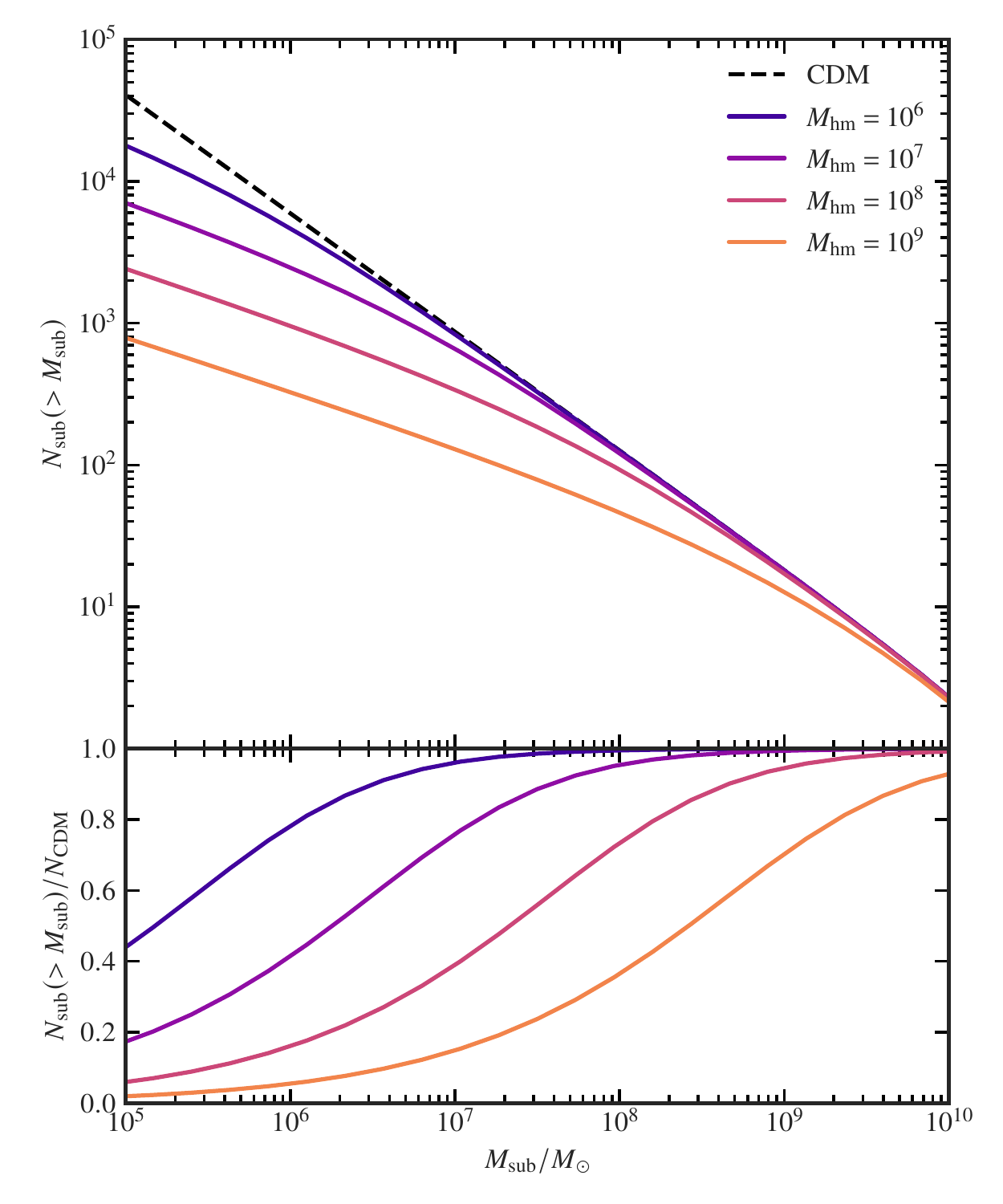}
	\caption{\label{fig:mass-function} Upper frame: the relative number of subhaloes $N_\mathrm{sub}$ with mass above $\Msub$ for the mass function in \cref{eq:mass-function}, at different half mode masses $\Mhm$. The dashed black curve is for a CDM model where $\Mhm=0$. In CDM, the number of haloes above a certain mass goes up by roughly one decade for every decade lower in mass. Lower frame: the number of subhaloes with mass above $\Msub$ relative to the same number in CDM, at the same half mode masses.}
\end{figure}

\begin{figure*}
	\includegraphics[width=1.0\textwidth]{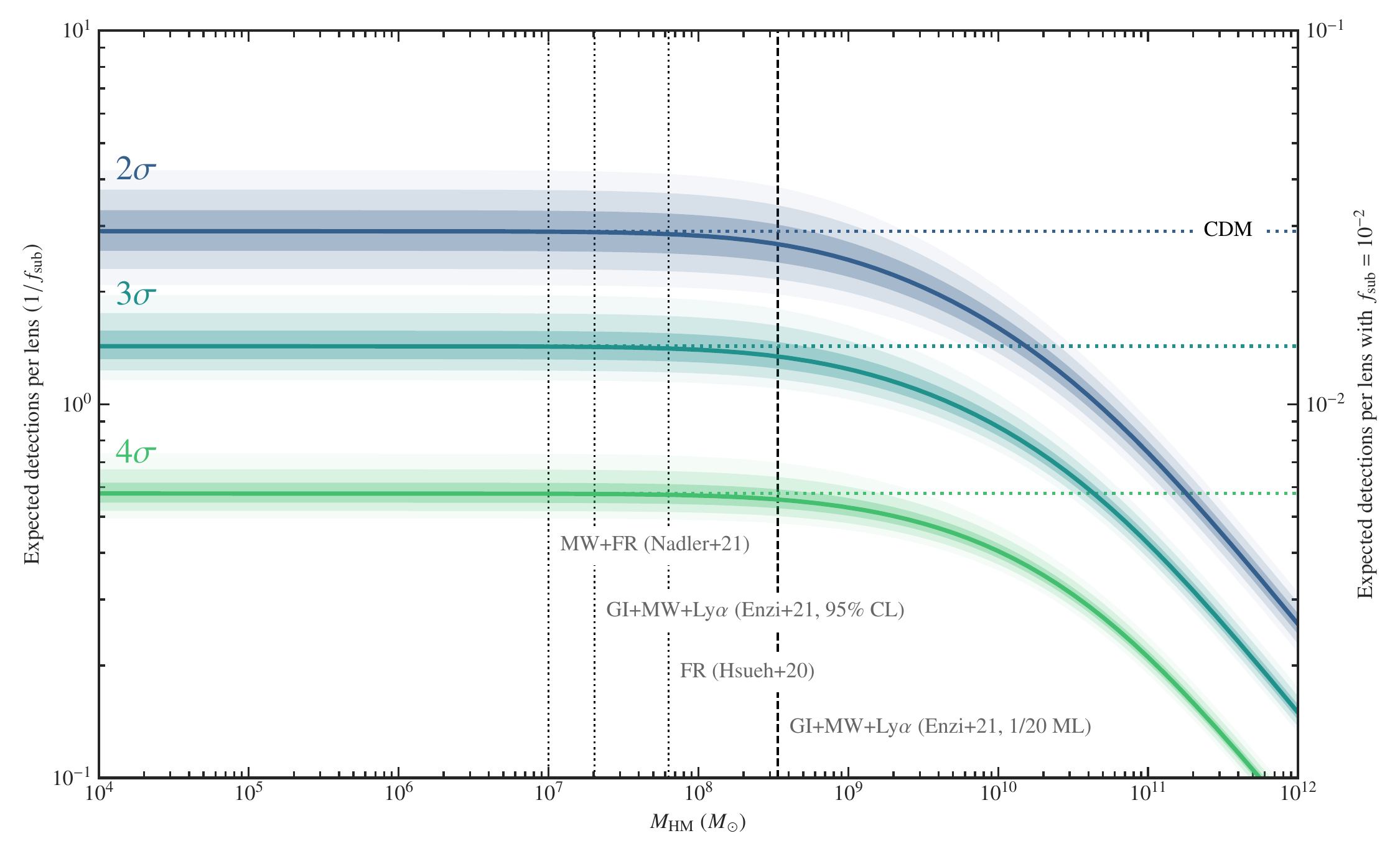}
	\caption{\label{fig:expected_number_mhm} Detectable subhaloes per lens, averaged over all lenses, as a function of half mode mass $\Mhm$. The left-hand axis gives the value per $\fsub$, the right-hand axis gives the value at $\fsub=0.01$. All values scale linearly with $\fsub$. Each curve is for subhalo detections at different significances, which are labelled. The $64\%$, $95\%$, and $99\%$ confidence areas are plotted with each curve. Horizontal dotted lines show the expected detectable subhaloes in CDM ($\Mhm\rightarrow0$). Vertical dotted lines show the current $95\%$ upper limits on $\Mhm$ from other studies, with the vertical dashed line showing the more conservative $1/20$th of the maximum likelihood from \citet{Enzi2021}.}
\end{figure*}

\begin{figure*}
	\includegraphics[width=1.0\textwidth]{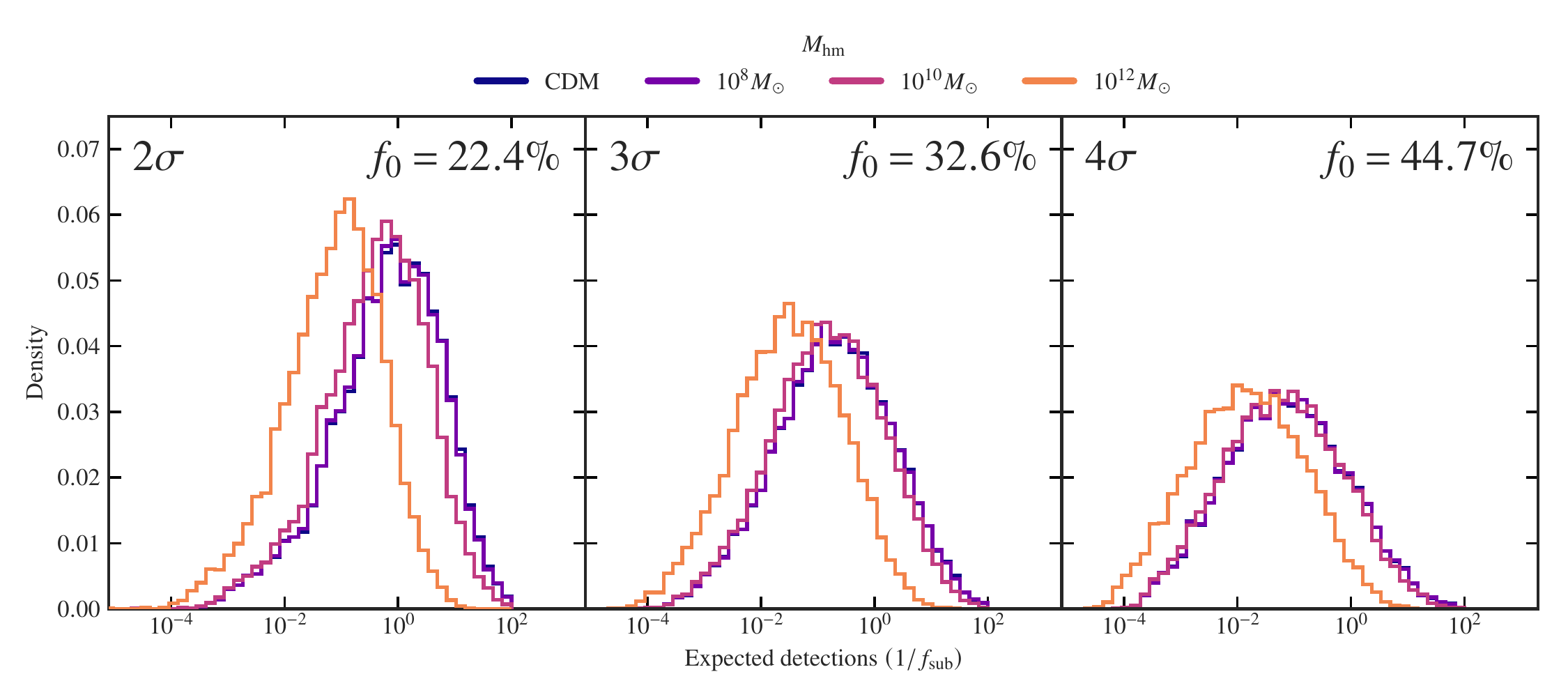}
	\caption{\label{fig:expected_number_significances} The distribution of the number of expected detections in each lens for different labelled detection significances, and different half mode masses, $\Mhm$. As in \cref{fig:expected_number_mhm}, the expected number is given in units of $1/\fsub$. In the top right of each frame the fraction of lenses for which the detectable number of haloes is zero, $f_0$, is given.}
\end{figure*}

In \cref{fig:sensitivity-statistics} we plot two distributions of sensitivity statistics computed from the sample. The upper frame plots the cumulative fraction of area inside $2\erad$ which is sensitive at the indicated mass. We find that $26.7$ per cent of the area in our simulated strong lens observations are sensitive enough to detect a subhalo with $\Mmax\leq 10^{11}\Msun$ at $3\sigma$. The same fraction is $2.35$ per cent for $\Mmax\leq 10^{10}\Msun$, and $0.03$ per cent for $\Mmax\leq 10^{9}\Msun$. The lower frame shows the distribution of sensitivity in each system's most sensitive pixel. With a $3\sigma$ detection threshold, the best pixel in the entire sample has a sensitivity of $\Mmax=10^{8.8\pm0.2}\Msun$. This represents a fundamental limit of subhalo detectability for this instrument.

\subsection{Mass function statistics}
\label{sec:detections}
Using the sensitivity maps, we can predict the expected number of detectable haloes in each lens for a given dark matter model. The number of subhaloes $\dd{n}$ of mass $m$ in a mass interval $\dd{m}$ per projected area on the sky is proportional to the subhalo mass function
\begin{equation}
	\label{eq:mass-function}
	\derivative{n}{m} \propto m^{\alpha_1}\left[1+\left(\alpha_2\frac{\Mhm}{m}\right)^\beta\right]^{\gamma},
\end{equation}
where the constants have the following values; $\alpha_1=-1.9$, $\alpha_2=1.1$, $\beta=1.0$, and $\gamma=-0.5$. The mass function comes from fitting to the data in \citet{Lovell2020b}, but the fit parameter values have been recalculated so that the mass function is in terms of $\Mmax$, rather than $M_\mathrm{sub}$. The half-mode mass, $\Mhm$, is the mass scale at which the square-root of the power-spectrum ratio between WDM and CDM is half. When $\Mhm=0$, the mass function is just that for CDM, i.e., $\dd{n}/\dd{m}\propto m^{\alpha_1}$. In \cref{fig:mass-function} we plot the mass function for different half mode masses. The expected number, $\musub$, of subhaloes in a mass range $\mmin'\leq m \leq \mmax'$ inside a projected radius $\theta=2\erad$ is given by
\begin{equation}
	\label{eq:mu-integral}
	\musub=\fsub^\mathrm{CDM}\Menc \frac{\displaystyle\int_{\mmin'}^{\mmax'}\dv{n}{m}\dd{m}}{\displaystyle\int_{\mmin}^{\mmax}m^{\alpha_1+1}\dd{m}}.
\end{equation}
where $\Menc$ is the mass of the lens inside that radius and $\fsub$ is the fraction of mass contained in substructure \citep{Vegetti2009b}. In this way $\fsub$ normalises the mass function.

For each pixel in each image we calculate $\musub$. We first choose a dark matter model, parametrised with $\Mhm$ and $\fsub$. We then evaluate the integrals in \cref{eq:mu-integral}, using the mass limits $m_0=10^6\Msun$ and $m_1=10^{11}\Msun$ in the denominator. In the numerator $m_1'$ is the same but the lower limit, $m_0'$ comes from the sensitivity map value for that pixel and the chosen detection significance. This lower mass limit has an associated uncertainty, as a result of the fitting procedure referred to in \cref{sec:sensitivity-estimation}. In order to account for this, we draw $10^4$ realisations of each sensitivity map from a Gaussian distribution centred on the fitted value of $m_0'$ with standard deviation given by the associated uncertainty. Multiplying by the ratio of pixel area to area inside $2\erad$ then gives the detectable $\musub$ in that pixel, at that significance, for that realisation. The expected number is then summed over all pixels to find a distribution of $\musub$ for each lens and dark matter model.

In \cref{fig:expected_number_mhm} we plot the expected number of detectable haloes per lens, averaged over our entire sample for a range of $\Mhm$. For a CDM universe we expect $\mu_\mathrm{sub}=1.43^{+0.14}_{-0.11}[\fsub^{-1}]$ subhaloes to be directly detectable per lens at $3\sigma$. This number scales linearly with $\fsub$, according to \cref{eq:mu-integral}. The expected number of detectable subhaloes is consistent with CDM for any value of $\Mhm$ below the current constraints on that parameter. Three such constraints are shown in the figure, each a $95\%$ upper limit. These come from: a combined analysis of the abundance and properties of Milky Way satellites and flux ratio anomalies in strong gravitational lenses \citep{Nadler2021b}; a combined analysis of gravitational imaging in strong lenses with extended sources, Milky Way satellites and the Lyman-$\alpha$ forest \citep{Enzi2021}; and, strong lensing flux ratio anomalies alone \citep{Hsueh2020}. Using the more conservative $1/20$ of the maximum likelihood, \citet{Enzi2021} also rule out models with $\Mhm>4.8\times10^8\Msun \mathrm{h}^{-1}$. To place competitive constraints on the half mode mass, the number of detections would need to be suppressed relative to CDM for $\Mhm<10^8$. The sensitivity limit we find in \cref{fig:sensitivity-statistics} therefore places the constraining power of strong lens images with the characteristics of Euclid VIS outside the region of interest for gravitational imaging.

\subsection{Substructure fraction}

The expected number depends crucially, but straightforwardly, on $\fsub$, which is as yet a poorly constrained quantity. \citet[][]{Despali2017} find $\fsub\approx 1\times10^{-2}$ in dark matter only simulations, and $\fsub\approx 5\times10^{-3}$ in the Illustris and EAGLE hydrodynamic simulations. \citet{Hsueh2020} measure $\fsub\approx 2\times10^{-2}$ in seven lensed quasars. These values have been converted to cover our larger mass range, but not our different definition of subhalo mass. In any case, both studies indicate that the order of magnitude of $\fsub$ is $10^{-2}$. For such an $\fsub$ our results predict that, at $86$ ($99$) per cent confidence, between one in $64$ $(51)$ and one in $76$ $(86)$ Euclid VIS lenses will yield a $3\sigma$ subhalo detection. This number is consistent with the results of gravitational imaging studies in HST data, where detections have been rare so far. \citet{Vegetti2014} detected one subhalo in $11$ lenses, and \citet{Nightingale2022} report only two convincing detections in $54$ lenses. Considering the superior resolution of HST versus VIS, and the selection of lenses in those studies relative to the broader sample used here, a smaller number of detections in Euclid VIS should be expected.

\subsection{Population statistics}

\begin{figure*}
	\includegraphics[width=1.0\textwidth]{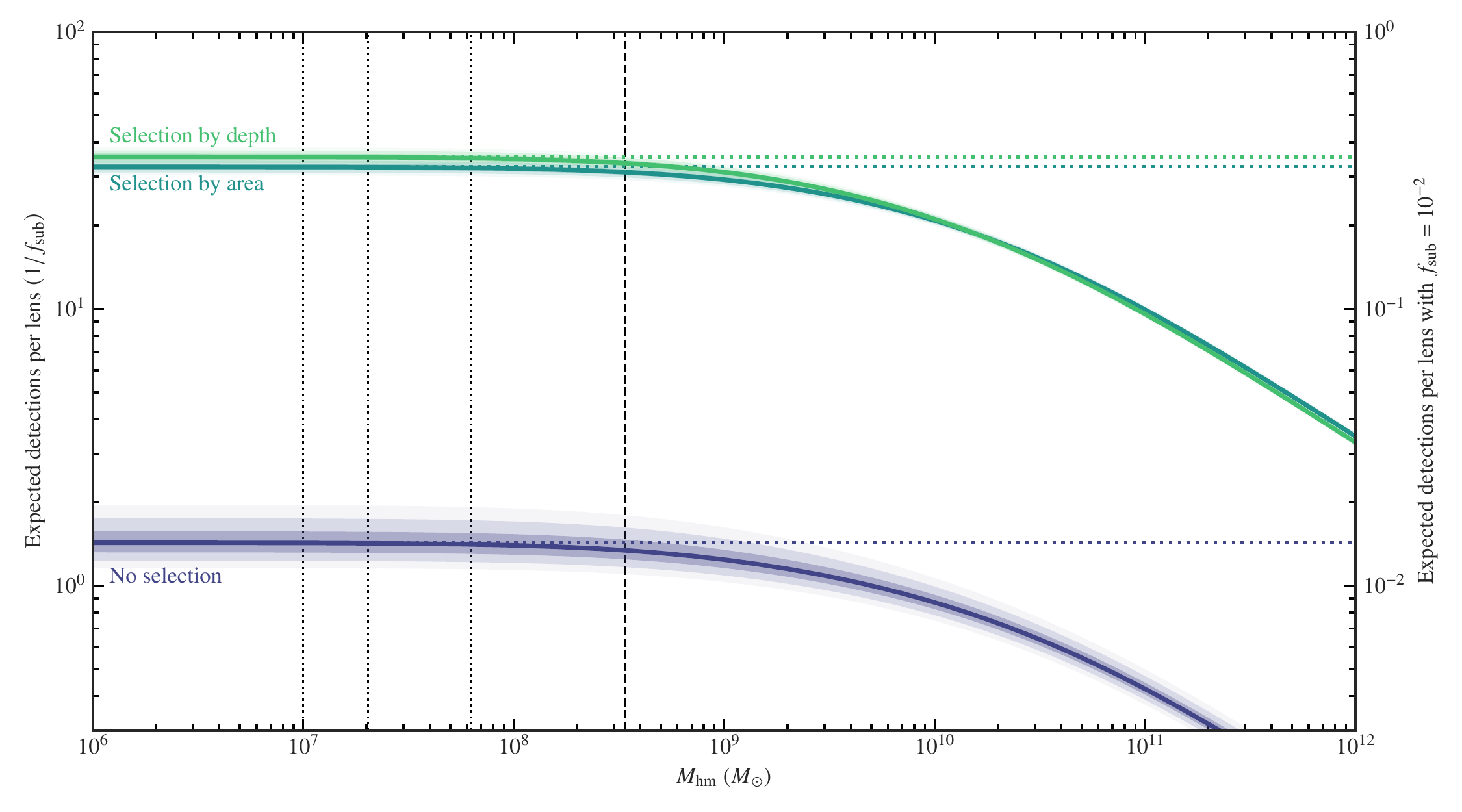}
	\caption{\label{fig:selection} Expected number of $3\sigma$ detectable haloes per lens as a function of half mode mass for: an average lens expected from the Euclid survey, the top 100 lenses with the largest sensitive area, and, the top 100 lenses with the deepest sensitivity. Horizontal dashed lines show the expected number in CDM. Vertical lines show the same upper limits as in \cref{fig:expected_number_mhm}.}
\end{figure*}

The number of expected detections differs greatly between strong lens systems. \cref{fig:expected_number_significances} plots the distribution of detections per system for different half mode masses and detection significances. The distribution is highly skewed towards the most sensitive systems. Specifically, $50$ per cent of all expected detections at $3\sigma$ in a CDM universe come from the top $2$ per cent of systems, and $90$ per cent of detections from the top $14$ per cent of systems. For $32.6$ per cent of systems no detections are expected at $3\sigma$ in CDM. The reason for this highly skewed distribution is obvious upon inspection of the mass function in \cref{fig:mass-function}. Every factor of $10$ improvement in sensitivity brings a factor of $10$ increase in the number of detectable subhaloes. This also accounts for the behaviour at different half mode masses. As $\Mhm$ increases, the most sensitive systems, i.e., those with the highest number of expected detections, are affected most strongly. This is because their expected detections are predominantly at lower masses, which are suppressed sooner than larger masses as $\Mhm$ increases.

\subsection{Selecting for sensitivity}
\label{sec:selection}

In the near future, analysis of strong lens images in large surveys will be used to constrain dark matter models. However, \cref{fig:expected_number_significances} shows that the constraining power for a majority of the images in the sample is weak or non-existent. This is a result of the fact that we have as yet never had such a large and homogeneous sample of strong lenses. According to \citetalias{Collett2015}, the majority of strong lenses we find in Euclid will have smaller Einstein radii and lower S/N than samples used for gravitational imaging in the past e.g. SLACS \citep{Bolton2006} or BELLS \citep{Brownstein2012}. The selection effects in these smaller samples produced lenses which were already relatively sensitive to substructure. Pre-selecting the most sensitive systems from a large sample which are mostly poor in sensitivity should drastically improve the constraints in a gravitational imaging study where analysis time is limited.

We propose two selection criteria. Sensitivity depth ranks systems by their most sensitive pixels, with the lowest mass being the best. This selection gives us systems at the low-mass end of the distribution in the lower frame of \cref{fig:sensitivity-statistics}. Sensitivity area ranks them by the total number of pixels which are sensitive at any mass. For both selections, we choose the best $100$ systems. These two selections are mass function independent, which is why we do not propose selecting for $\mu$ directly.

In \cref{fig:selection} we show the effect of pre-selection on the expected number of detections at $3\sigma$. With selection, the expected number of $3\sigma$ detections in CDM increases to $32.6^{+0.8}_{-0.8}[\fsub^{-1}]$ and $35.6^{+0.9}_{-0.9}[\fsub^{-1}]$ per lens for selection by area and by depth respectively. Using our fiducial $\fsub=10^{-2}$, pre-selection then gives one detection in every $\sim$ three lenses, up from one in $\sim70$ with no selection. The limit of sensitivity cannot improve with selection, so we do not expect constraints on $\Mhm$ to change. However, the prospects for constraining $\fsub$ improve dramatically. 

\begin{figure}
	\includegraphics[width=1.\columnwidth]{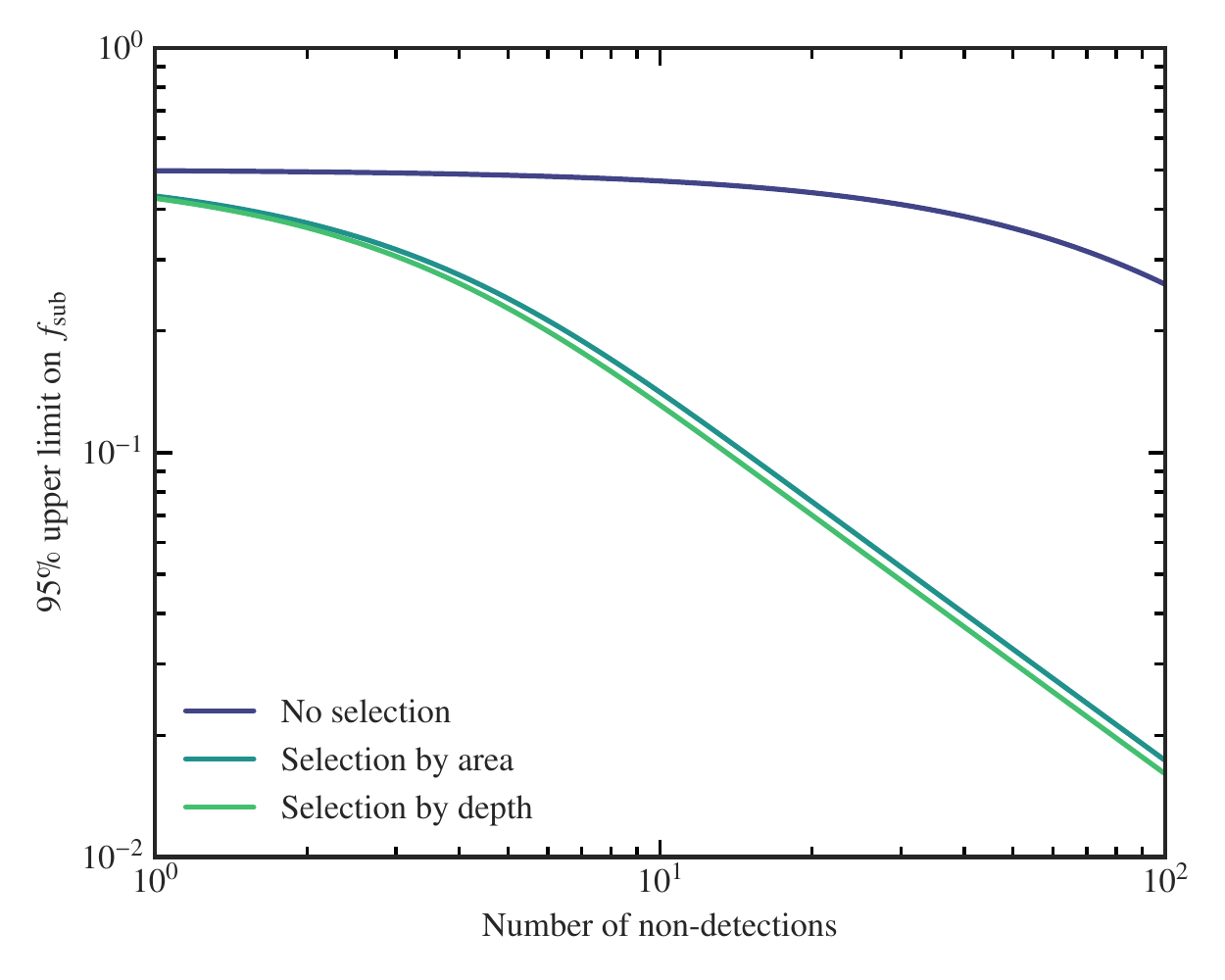}
	\caption{\label{fig:selection-constraints} The $95\%$ upper limit on $\fsub$ from $N$ non-detections in data sets of systems with no pre-selection, and with pre-selection by sensitivity area and depth. These data assume a CDM mass function ($\Mhm=0$) such that $\musub$ only depends on $\fsub$.}
\end{figure}

We illustrate this in \cref{fig:selection-constraints} which shows the possible constraints on $\fsub$ in a dataset of non-detections where, before analysis for subhaloes, the lenses are selected randomly versus the selection methods we described previously. Assuming CDM, we plot the $95\%$ upper limit on $\fsub$ from $N$ non-detections in data sets using no selection and the two selection methods described previously. It is important to note that non-detections are only informative in this case because $\musub$ for the three different sets of lenses has been calculated from the sensitivity maps. The probability of a non-detection in the mass range which the lens is sensitive to can then be calculated with Poisson statistics. The figure shows that $10$ non-detections in an analysis of $10$ random Euclid strong lenses can place an upper limit on $\fsub$ of only $0.46$, because the probability of non-detections in those lenses is already very high. With selection, this decreases to $0.14$ with either method. With $100$ non-detections the same limits are $0.26$ without selection and $0.016$ with selection. The latter is well within the region of interest for $\fsub$ indicated from simulations. Considering also that any detection in this sample immediately gives a strict lower limit on $\fsub$, it is clear that a reasonable number of highly sensitive Euclid strong lenses can place useful constraints on the substructure mass fraction. With such a large number of lenses available $\fsub$ could also be characterised as a function of galaxy environment and redshift, providing useful data for galaxy evolution studies.

\section{Discussion}
\label{sec:discussion}

\subsection{Limits of training data}

As with any machine learning method, the extent to which our results are useful depends on the realism of the training data. In our case, the question is whether there are shortcomings in the training data which make the neural network's subhalo detection task easier relative to that in real data. Here we discuss three possible extensions and their likely effect on the results.

Our choice of Hubble Deep Field sources is intended to reflect the complexity of real lensed sources and account for the degeneracy between source brightness and dark substructure which is often neglected in other similar work. However, in a traditional forward modelling approach, there is no imposed constraint that the source must be realistic. Typically, a regularisation condition is imposed on a pixelated source such that is must be locally smooth but as long as this is satisfied the shape can be arbitrarily complex. Addressing this would require training a new model with sources of arbitrary shape and complexity and comparing the results in terms of sensitivity. This space of all possible sources is not well defined which makes performing this test difficult.

In our mock observations the lens light surface brightness is always a perfect S\'ersic profile which is also perfectly subtracted, leaving only the Poisson noise from the very bright centre of the lens in the mock observation. See, for example, the noise in the centre of the lower left image in \cref{fig:sensitivity-maps}. This choice was motivated by simplicity and computation speed when generating many millions of images. Lens galaxies, being almost always ellipticals, are often well described by S\'ersic profiles. However, any non-S\'ersic component in the lens light leaves residual surface brightness which can be degenerate with the effect of the subhalo. We expect that a network trained with imperfect subtractions can learn to take this degeneracy into account, as our current model did with other degeneracies discussed in \cref{sec:model-performance}.

For the lens galaxy mass profile, our choice of an elliptical power-law also simplifies the problem slightly for the network, relative to reality. If the lens mass model cannot absorb smaller, local changes in mass then detecting substructure becomes easier and sensitivity improves. In strong lensing studies a power-law, even with a fixed isothermal slope, is typically sufficient to fit the positions and fluxes of the lensed images. For gravitational imaging, more complexity in the mass model is needed to avoid false positive subhalo detections. Adding multipole, disc, or extra power-law components to the lens mass would complicate its structure and make subhaloes in certain positions relative to the lens harder to detect. The exact nature of the degeneracies between lens mass and light models, and the presence of substructure will be the subject of future papers.

For the three problems just discussed, we expect that extensions of the training data to address them would degrade model performance specifically because they introduce effects which are degenerate with the subhalo signal. They can make it harder to detect subhaloes in specific systems where these effects might be strong, but, they cannot change the fundamental sensitivity limit that we observe in \cref{fig:sensitivity-statistics} which is ultimately set by the instrument resolution and seeing. We therefore expect that, with these extensions, the total expected number of detectable subhaloes could decrease, but, the dependence of this number on the half mode mass would not change.

\subsection{Subhalo concentration}

The concentration of a subhalo has a significant effect on its detectability \citep{Amorisco2022}. As the concentration of a subhalo's mass profile increases, its effect on the lensed images becomes more localised. This makes it easier to differentiate the signal of the subhalo from that of other sources which tend to have smoother effects, e.g. from the lens mass macro model, the lens light subtraction, or from other perturbing objects in the field of view. When the subhalo is less concentrated, the opposite is true, making them harder to detect.

To produce the results in \cref{sec:results}, we used the concentration from \cref{eq:rmax-vmax-relation} which is that for CDM. In simulations, subhaloes in warmer dark matter models  are found to be less concentrated relative to CDM. In theory, subhaloes in warmer models are then harder to detect. This causes a further suppression in the expected number of detections at large values of $\Mhm$. If this suppression is significant relative to that from the mass function, then the difference between CDM and WDM in e.g. \cref{fig:expected_number_mhm} may be more exaggerated. This in turn allows for stronger constraints on $\Mhm$.

\begin{figure}
	\includegraphics[width=\columnwidth]{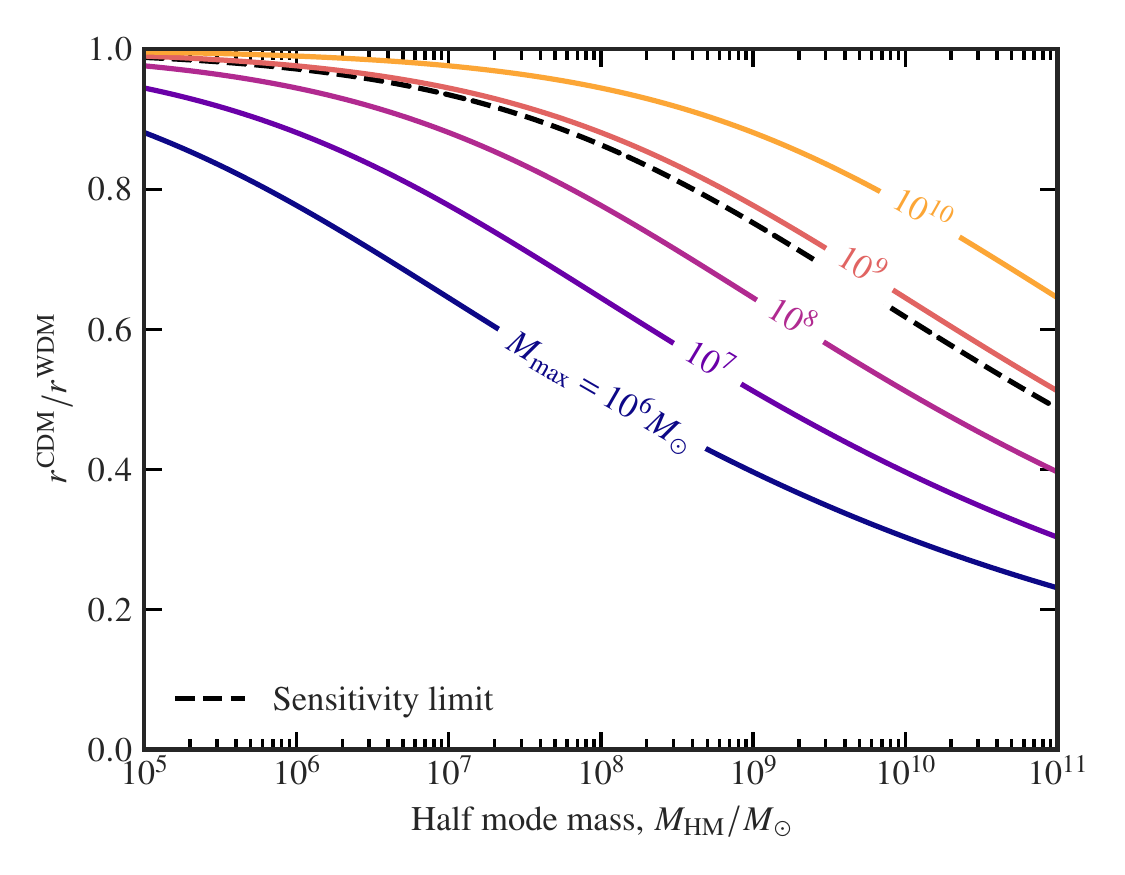}
	\caption{\label{fig:concentration-correction} The change in subhalo concentration in warm dark matter models relative to that in CDM. Each curve plots the change in concentration for different values of $\Msub$, which is labelled. The dashed black curve is the sensitivity limit in subhalo mass which we found in \cref{sec:results}. All subhaloes detectable at $3\sigma$ in our results are above this mass. An increase (decrease) in $\rmax$ relative to CDM corresponds to a decrease (increase) in concentration.}
\end{figure}

The concentration is parametrised by the subhalo's $\rmax$, and the correction to $\rmax$ in WDM relative to CDM is given by
\begin{equation}
	\label{eq:concentration-correction}
	\frac{\rmax^\mathrm{CDM}}{\rmax^\mathrm{WDM}}=\left[1+\alpha \left(\frac{\Mhm}{\Mmax}\right)^\beta\right]^\gamma,
\end{equation}
where $\alpha$, $\beta$, and $\gamma$ are constants derived from simulations with the values $2.0$, $0.4$ and $-0.3$ respectively. This fit is obtained using the data sets discussed in \citet{Lovell2020b} together with an adaptation of their method, as applied to the halo mass-$\rmax$ instead of the halo mass function. We consider six pairs of hydrodynamical simulation data sets that describe WDM and CDM, with values of $\Mhm$ for the WDM model that span the range of $\Mhm=[1.3\times10^8,3.5\times10^9]\Msun$. For the subhaloes of each data set we compute the median $\rmax$ as a function of $\Mmax$ and then calculate the ratio of the counterpart CDM and WDM simulations' median relations. We then perform a simultaneous fit to all six data sets using the functional form presented in \cref{eq:concentration-correction}, and obtain the parameter values discussed above.  We have also performed this procedure for isolated haloes, using the $M_{200}$ mass definition in place of $\Mmax$, and in that case obtain $\alpha=4.0$, $\beta=0.3$, and $\gamma=-0.6$. Note that both mass-$\rmax$ ratio fits place CDM in the numerator and WDM in the denominator: this choice is made because $\rmax$ increases with $\Mhm$. We plot the change in concentration in \cref{fig:concentration-correction}.

It is not clear if such a change in concentration is large enough to effect subhalo detectability. As such, we test this effect explicitly by computing sensitivity maps at $10$ different half mode masses for a small number of systems with concentrations set now by \cref{eq:concentration-correction}, rather than the CDM concentration used previously. This is possible without retraining our original neural network because the subhaloes in the training data did not use any concentration-mass relation. Rather, $\rmax$ was drawn from a log-uniform distribution spanning more than the range to be tested here. Computing new sensitivity maps with many values of $\Mhm$ is a prohibitively expensive task, so we choose to use the $100$ best systems selected for sensitivity depth in \cref{sec:selection}. These systems have the best sensitivity in the region where the concentration correction is strongest, and so, if the effect of concentration is significant, it will be apparent in these systems more readily than others.

\begin{figure}
	\includegraphics[width=1.0\columnwidth]{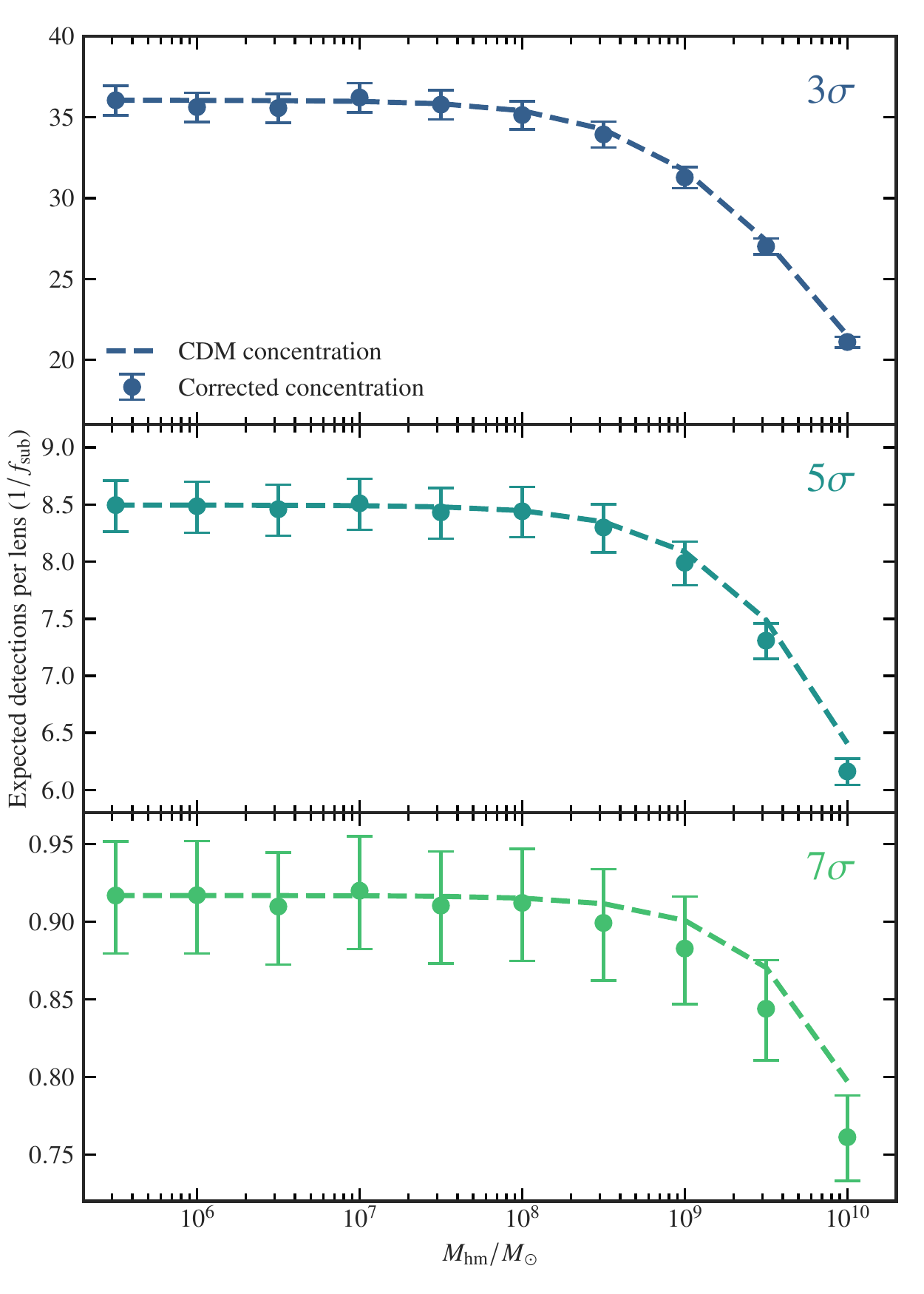}
	\caption{\label{fig:concentration-expected-numbers} The expected number of detectable subhaloes at $3\sigma$, $5\sigma$, and $7\sigma$ for a sample of $100$ high sensitivity lenses using a fixed CDM subhalo concentration (dashed line) and a concentration corrected for WDM (points). The $1\sigma$ error bars are plotted with the corrected concentration points.}
\end{figure}

In \cref{fig:concentration-expected-numbers} we plot the number of detectable subhaloes in these systems for the corrected and uncorrected sensitivity functions. For the relatively low detection threshold of $3\sigma$ used in our earlier results, we find a change in the expected number of detections consistent with zero. This changes as we increase the detection threshold. At the largest $\Mhm$ tested ($10^{10}\Msun$), the suppression in the number of detections relative to the CDM concentration is $(4\pm2)$~per~cent at $5\sigma$, and $(5\pm4)$~per~cent at $7\sigma$. As a subhalo's concentration strongly effects its detectability, it follows that for higher detection thresholds, concentration has a stronger effect. At the detection thresholds typically used for gravitational imaging studies, i.e. $\sim 10\sigma$ or equivalent, we expect the effect to be stronger still, as others have already shown.

We should also note that for the subhaloes considered here, the actual change in concentration in warmer dark matter models is relatively small. Consider the region above the dashed curve in \cref{fig:concentration-correction}. This area covers all the detectable subhaloes in Euclid images, in the `best-case' concentration scenario, CDM. A $10^{10}\Msun$ subhalo only undergoes a change in concentration of $\sim20$ per cent at the largest $\Mhm$ tested, and changes by only a few per cent in the region of interest for $\Mhm \leq 10^{8}\Msun$. Subhaloes where the WDM concentration has a significant effect are already undetectable in CDM, especially at the higher detection thresholds where concentration matters more. The suppression due to concentration changes should therefore not significantly effect the ability of Euclid images to constrain $\Mhm$.

\subsection{Field haloes}

In this work we only considered subhaloes, i.e., haloes inside the main lensing galaxy's halo. For these haloes it is sufficient to treat them as being in the same plane as the lensing galaxy. However, numerical simulations show that small haloes also exist in the field, separated from any galaxy-sized halo. \citet[][see their table 3]{Despali2018} give estimates for the number density of field haloes in the volume along the line of sight between observer and source, for different lens and source redshifts. For subhalo masses and lens and source redshifts probed by Euclid, the number of field haloes is much greater than the number of subhaloes. However, including these field haloes does not necessarily boost the number of detections. This is because objects away from the plane of the lens are harder to detect. \citet{Amorisco2022} give sensitivity maps for field haloes placed at different redshifts in front of and behind the lens. Differences in method and definitions mean we cannot directly compare with these maps but the general result is still useful. In all cases the sensitivity for these objects degrades relative to subhaloes in the lens plane.

There are two implications for our own results. First, field haloes are more numerous than subhaloes but in general are harder to detect. The increase in expected number is larger than the decrease in detectability so the net effect would be to boost the number of detections in general. However, this increase does not improve possible constraints on $\fsub$, as this quantity is only related to the fraction of mass in substructure in the lens galaxy halo. Second, according to \citet{Amorisco2022} there is no situation where a field halo is more detectable than its equivalent substructure and so the fundamental sensitivity limit we found in \cref{sec:results} would not improve. The addition of field haloes to our work would therefore not improve the prospects for constraining the half mode mass with Euclid VIS images, but their detection will provide constraints on the subhalo mass function's normalisation.

\section{Conclusions}
\label{sec:conclusions}

We have developed a machine learning based method for estimating sensitivity to dark substructures in strong lenses. We specifically targeted the Euclid survey and its VIS instrument, as this will provide the largest single sample of strong lens images to date. Our convolutional neural network is trained to detect dark matter subhaloes in mock images with: elliptical power-law lenses, sources from the Hubble Deep Field, external shear, and noise and PSF resembling Euclid VIS. Our neural network successfully learned some of the degeneracies present in traditional strong lens modelling, although it is not explicitly trained to do so. For example, it learns that the effect of a subhalo in the centre of the lens is degenerate with the lens mass model, where an increase in lens mass can produce the same effect. Sensitivity maps accordingly show no sensitivity to subhaloes in the centre of the lens.

We simulated $16$,$000$ strong lens images with the resolution and S/N of Euclid VIS, and realistic parameter distributions, modelled after \citet{Collett2015}. We then used our trained network to estimate the subhalo sensitivity in every image. Assuming a $3\sigma$ subhalo detection threshold, we found that $2.35$ per cent of pixels inside twice the Einstein radius were sensitive to subhaloes with a mass $\Mmax\leq 10^{10}\Msun$, $0.03$ per cent were sensitive to a mass $\Mmax\leq 10^{9}\Msun$, and, the limit of sensitivity in the instrument was found to be $\Mmax=10^{8.8\pm0.2}\Msun$.

From the generated sensitivity maps we were also able to predict the number of detectable subhaloes per lens, given a dark matter model and subhalo mass function. In CDM, we expect $\mu_\mathrm{sub}=1.43^{+0.14}_{-0.11}[\fsub^{-1}]$ detectable subhaloes per strong lens imaged in a Euclid-like survey. Assuming a substructure mass fraction of $\fsub=0.01$, this gives a detectable $3\sigma$ subhalo in one in every $\sim70$ lenses. This low number reflects the diversity and magnitude of the Euclid strong lens sample. If one selects only the best lenses in terms of sensitivity, the expected number of detections increases to $35.6^{+0.9}_{-0.9}[\fsub^{-1}]$ per lens for the $100$ most sensitive lenses. Again assuming $\fsub=0.01$, this gives one detectable subhalo in every $\sim$~three lenses. With selection, gravitational imaging in images like those of Euclid VIS can therefore give useful constraints on the substructure mass fraction $\fsub$. We show for example that $100$ non-detections in the most sensitive lenses would give $\fsub<0.016.$, an upper limit close to estimates from simulations. If Euclid indeed finds $\sim170$,$000$ new strong lenses as predicted, the number of subhalo detections should number in the thousands.

We also find that the expected number of detectable subhaloes does not change relative to CDM in WDM models which have not already been ruled out. This is a consequence of the sensitivity limit we find at $10^{8.8\pm0.2}\Msun$, for a $3\sigma$ detection. A number of methods have already placed upper limits on $\Mhm$ below this mass. This limit is primarily a function of the instrument pixel scale and seeing, and is higher in mass than the typical sensitivity of HST images which have a pixel scale roughly half as small.

Finally, we consider the suppression in the number of detectable subhaloes due to reduced subhalo concentration in warmer dark matter models. At the $3\sigma$ detection threshold used for our main results, we find no suppression in $\musub$ due to reduced concentration. The actual change in concentration for the subhalo masses that our images are sensitive to is relatively small. However, at detection thresholds of $5\sigma$ and $7\sigma$ we find a small suppression. At higher thresholds therefore, a subhalo needs to be more concentrated to remain detectable.

\section*{Acknowledgements}

CO'R thanks the members of the Euclid Strong Lens Science Working Group (SL-SWG) for useful feedback and discussion, as well as the Max Planck Computing and Data Facility (MPCDF) for computational resources and support. CO'R also thanks Benjamin Holzschuh for useful insight on the machine learning aspects of this work. GD was supported by a Gliese Fellowship. SV thanks the Max Planck Society for support through a Max Planck Lise Meitner Group. SV acknowledges funding from the European Research Council (ERC) under the European Union's Horizon 2020 research and innovation programme (LEDA: grant agreement No 758853). MRL acknowledges support by a Grant of Excellence from the Icelandic Research Fund (grant number 206930).

\section*{Data Availability}
The data used in this paper are available from the corresponding author on request.

\bibliographystyle{mnras}
\bibliography{bibliography}

\bsp	
\label{lastpage}
\end{document}